\documentclass[lettersize,journal]{IEEEtran}
\usepackage{cite}
\usepackage{amsmath,amssymb,amsfonts}
\usepackage{algorithm}
\usepackage{algpseudocode}
\usepackage{graphicx}
\usepackage{textcomp}
\usepackage{xcolor}
\usepackage{enumitem} 
\graphicspath{ {images/} }
\usepackage{makecell} % 加载 makecell 宏包

\begin{document}
	
	\title{Joint Contact Planning for Navigation and Communication in GNSS–Libration Point Systems}
	
	\author{
		Huan Yan,
		Juan A. Fraire,~\IEEEmembership{Member,~IEEE,}
		Ziqi Yang,
		Kanglian Zhao,~\IEEEmembership{Member,~IEEE,}
		Wenfeng Li,~\IEEEmembership{Member,~IEEE,}
		Xiyun Hou,~\IEEEmembership{Member,~IEEE,}
		Haohan Li,
		Yuxuan Miao,
		Jinjun Zheng,
		Chengbin Kang,
		Huichao Zhou,
		Xinuo Chang,
		Lu Wang,
		Linshan Xue,
		
		% <-this % stops a space
		\thanks{H. Yan, Z. Yang, K. Zhao, W. Li, X. Hou, H. Li and Y. Miao are with Nanjing University, Nanjing 210008, China, E-mail: (yanhuan@smail.nju.edu.cn; 652022230035@smail.nju.edu.cn; zhaokanglian@nju.edu.cn; leewf\_cn@hotmail.com; houxiyun@nju.edu.cn; lihh@smail.nju.edu.cn; dz21260005@smail.nju.edu.cn); J. A. Fraire is with the CONICET, Instituto de Estudios Avanzados en Ingeniería y Tecnología (IDIT), Córdoba 5000, Argentina, and also with Saarland University, Saarbrücken 66123, Germany, E-mail: (juanfraire@unc.edu.ar); J. Zheng, C. Kang, H. Zhou, X. Chang, L. Wang, L. Xue are with the China Academy of Space Technology, Beijing 100094, China, E-mail: (zhjinjun@vip.sina.com; 75012565@qq.com; zhc202412@163.com; chang\_xinuo@foxmail.com; roadlu\_wang@qq.com; 201811040822@std.uestc.edu.cn). (Corresponding authors: Jinjun Zheng; Kanglian Zhao.)}
		
		% <-this % stops a space
		%\thanks{Manuscript received April 19, 2021; revised August 16, 2021.}
	}
	
	% The paper headers
	%\markboth{Journal of \LaTeX\ Class Files,~Vol.~14, No.~8, August~2021}%
	%{Shell \MakeLowercase{\textit{et al.}}: A Sample Article Using IEEEtran.cls for IEEE Journals}
	
	%\IEEEpubid{}
	% Remember, if you use this you must call \IEEEpubidadjcol in the second
	% column for its text to clear the IEEEpubid mark.
	
	\maketitle
	
	\begin{abstract}
		% Deploying satellites at Earth-Moon libration points can remedy the limitations of GNSS in deep space scenarios due to its low orbital altitude.
		% Combining LP constellations with GNSS into a joint constellation helps form a more robust, comprehensive PNT (Positioning, Navigation, and Timing) system;   
		% The joint constellation can also provide navigation and communication services to spacecraft (i.e., users) within cislunar space.
		% However, the substantial propagation delays arising from the long distances between LP satellites/users and GNSS satellites lead to differing link durations compared to those within the GNSS constellation.
		Deploying satellites at Earth–Moon Libration Points (LPs) addresses the inherent deep-space coverage gaps of low-altitude GNSS constellations.
		Integrating LP satellites with GNSS into a joint constellation enables a more robust and comprehensive Positioning, Navigation, and Timing (PNT) system, while also extending navigation and communication services to spacecraft operating in cislunar space (i.e., users).
		However, the long propagation delays between LP satellites, users, and GNSS satellites result in significantly different link durations compared to those within the GNSS constellation.
		% Scheduling inter-satellite links (ISLs) falls under the domain of Contact Plan Design (CPD). 
		% Prior CPD schemes focus solely on scheduling ISLs within GNSS, featuring a uniform link duration, failing to address the scheduling of ISLs with heterogeneous link duration units between different entities in a joint GNSS-LP constellation. 
		% To address this challenge, we propose a Joint CPD (J-CPD) scheme. 
		Scheduling inter-satellite links (ISLs) is a core task of Contact Plan Design (CPD).
		Existing CPD approaches focus exclusively on GNSS constellations, assuming uniform link durations, and thus cannot accommodate the heterogeneous link timescales present in a joint GNSS–LP system.
		To overcome this limitation, we introduce a Joint CPD (J-CPD) scheme tailored to handle ISLs with differing duration units across integrated constellations.
		The key contributions of J-CPD are: 
		\textbf{(i)} introduction of LongSlots (Earth-Moon scale links) and ShortSlots (GNSS-scale links);
		\textbf{(ii)} a hierarchical and crossed CPD process for scheduling LongSlots and ShortSlots ISLs;
		\textbf{(iii)} an energy-driven link scheduling algorithm adapted to the CPD process.
		%
		% Simulation results on a joint BeiDou–LP scenario show that J-CPD outperforms the baseline FCP method in delay and ranging coverage while keeping suitable user satisfaction and offering adjustable trade-offs via tunable energy parameters. 
		% This is the first CPD framework to jointly optimize navigation and communication services in GNSS–LP systems, marking a foundational step toward unified, resilient deep space PNT architectures.
		Simulations on a joint BeiDou–LP constellation demonstrate that J-CPD surpasses the baseline FCP method in both delay and ranging coverage, while maintaining high user satisfaction and enabling tunable trade-offs through adjustable potential-energy parameters.
		To our knowledge, this is the first CPD framework to jointly optimize navigation and communication in GNSS–LP systems, representing a key step toward unified and resilient deep-space PNT architectures.
	\end{abstract}
	
	
	\section{Introduction}
	\label{sec_introduction}
	\IEEEPARstart{G}{NSS} is a spatiotemporal infrastructure composed of multiple satellites that provide Positioning, Navigation, and Timing (PNT) services~\cite{38,39}. 
	There are four mature GNSS systems: the American GPS, Russian GLONASS, European Galileo, and Chinese BeiDou. 
	These GNSS systems form the cornerstone of the interconnected world in today's information society.
	
	GNSS has introduced inter-satellite links (ISLs) to reduce dependency on ground stations (GSs). 
	On the one hand, through GS-satellite links and ISLs, it is possible to manage satellites beyond GS visibility. 
	On the other hand, ISLs enable inter-satellite ranging and communication, enhancing the accuracy of satellite orbit determination and clock synchronization~\cite{40}.
	
	\begin{figure}[] 
		\centering
		\includegraphics[width=0.9\linewidth]{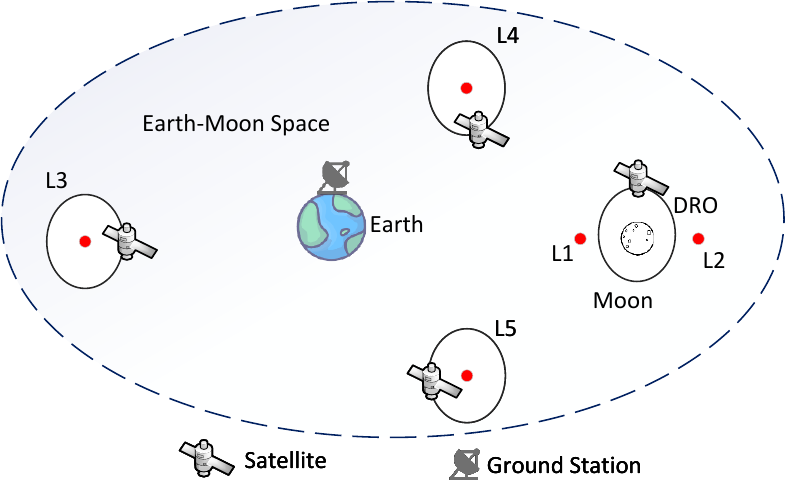}
		\caption{Illustration of a Libration Point (LP) constellation}
		\label{fig-1}
	\end{figure}
	
	However, limited by orbital altitude, the application of GNSS in deep space scenarios is restricted~\cite{41}. 
	Based on this fact, the concept of Comprehensive PNT has emerged~\cite{42}. 
	Comprehensive PNT aims to construct a multi-source PNT infrastructure based on different principles that can extend the scope of PNT services. 
	Libration points (LPs) are unique equilibrium points formed under the joint gravitational influence of two massive celestial bodies, such as the Moon and the Earth, in celestial mechanics. 
	There are five LPs in total, namely L1 to L5. Satellites can maintain a stable relative position at these LPs without frequent orbit adjustments. 
	Deploying navigation constellations at LPs is an ideal solution for constructing deep space time-space infrastructure and constitutes a crucial component of the comprehensive PNT system~\cite{7,8,9,10,17}. 
	Fig~\ref{fig-1} shows a typical LP constellation deployed at the Earth-Moon L3, L4, and L5 points and in the lunar Distant Retrograde Orbit (DRO).
	
	Considering the limitations of satellite payload, GNSS satellites typically carry a single ISL terminal. 
	GNSS satellites need to establish ranging links with as many other satellites as possible to achieve high-precision orbit determination and time synchronization. 
	This necessitates that the ISL terminals on GNSS satellites frequently switch their link partners within a short period. 
	Using a polling time division duplex (PTDD), radio frequency phased array narrow beams can quickly change beam direction, making it a feasible solution for ISL terminals~\cite{30,37}.
	
	\begin{figure}[] 
		\centering
		\includegraphics[width=0.9\linewidth]{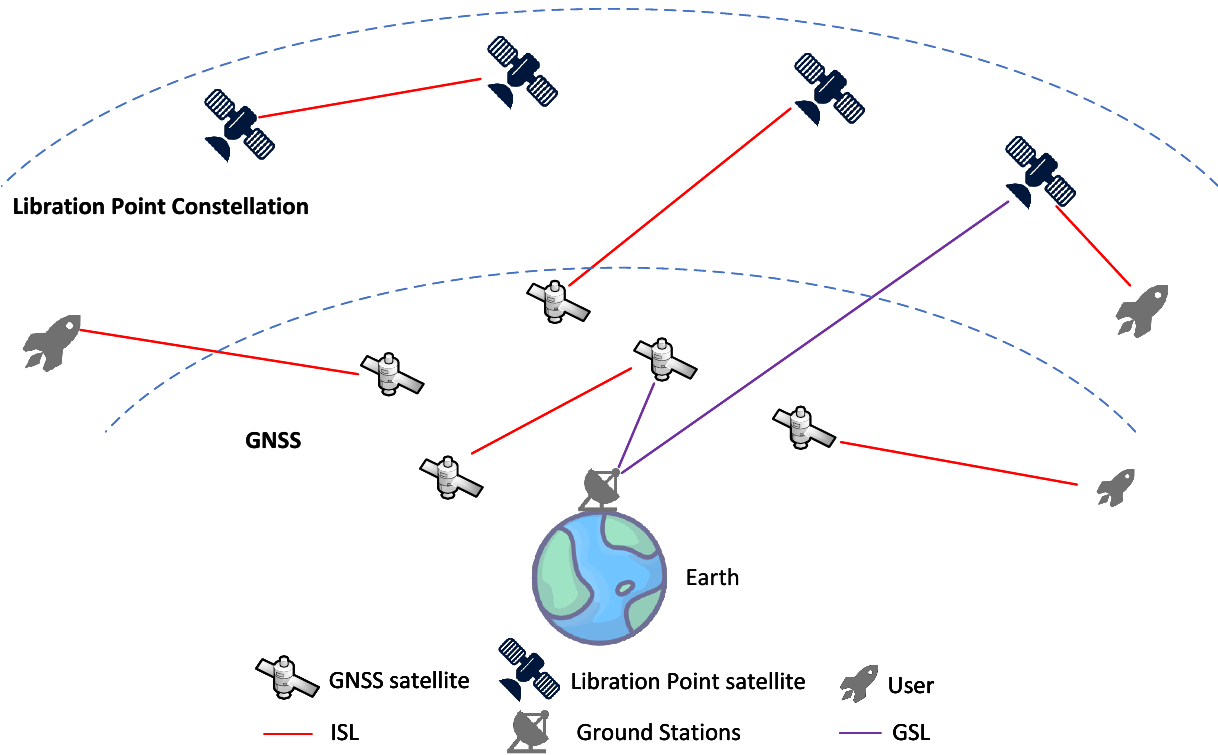}
		\caption{GNSS and LP joint constellation}
		\label{fig0}
	\end{figure}
	
	This paper focuses on GNSS and LP satellites that have a single RF-phased array narrow-beam terminal. 
	This setup establishes links between LP satellites and GNSS satellites, thereby giving rise to a joint constellation of GNSS and LP constellations as illustrated in Fig~\ref{fig0}. 
	We define the \textbf{joint constellation} as follows:
	\begin{enumerate}
		\item Mutual ranging between GNSS satellites and LP satellites is conducted, with the GNSS and the LP constellation functioning as a unified constellation to achieve precise orbit determination and time synchronization.
		\item Communication between GNSS satellites and satellites at LPs is feasible, establishing the GNSS and the LP constellation as an interconnected information network.
		\item Both the GNSS and LP constellations rely on common ground stations. These ground stations manage and control both constellations as a single integrated system.
		\item The GNSS constellation and the LP constellation serve space users collectively. In this paper, space users refer to external spacecraft running in cislunar space that require ISLs with satellites to obtain communication or navigation services.
	\end{enumerate}
	
	Constructing a joint constellation offers several advantages:
	\begin{enumerate}
		\item In a joint constellation, GNSS and LP satellites rely on the same ground stations, reducing the cost of establishing new ground infrastructure.
		\item Compared with ranging conducted solely within the GNSS constellation (or the LP constellation), GNSS satellites and LP satellites present novel geometric positions relative to each other. This indicates that ranging between GNSS and LP satellites constitutes a more diversified geometric structure, facilitating lower PDOP values~\cite{44} and enhancing orbit determination accuracy.
		\item Considering the GNSS and the LP constellation as an integrated system facilitates unified link resource allocation, thereby better serving space users.
		\item Analogous to BeiDou's global short message service~\cite{40}, a joint constellation can establish end-to-end communication methods at the Earth-Moon scale (e.g., enabling communications between handheld terminals on Earth and those on the Moon).
	\end{enumerate}
	
	However, scheduling the ISLs within a joint constellation becomes significantly more complex than that for GNSS. 
	Constructing an ISL scheduling scheme is called Contact Plan Design (CPD)~\cite{18}. 
	Existing GNSS CPD schemes~\cite{23,24,26,20,22,29,25,27,32,19,28,30} have the following limitations within the LP-GNSS joint constellation:
	\begin{enumerate}
		\item Propagation delay from the large distances between LP satellites/users and GNSS satellites causes a longer link duration unit compared to that between GNSS satellites. 
		Conventional GNSS CPD schemes only schedule ISLs with a single, uniform link duration within GNSS, while being incapable of handling the scheduling of ISLs that exhibit multiple different link durations within the LP-GNSS joint constellation.
		\item Link time costs between GNSS and LP satellites/users are higher than between GNSS satellites. Prior CPD schemes schedule only intra-GNSS ISLs with uniform time costs to meet the ranging/communication needs of GNSS satellites. However, in LP-GNSS constellations, CPD schemes must satisfy both LP satellite and GNSS satellite ranging/communication needs while better serving users, accounting for these varying time costs between GNSS satellites and diverse nodes.
		Prior GNSS CPD schemes fail to model the resource consumption impact of high time cost ISLs (e.g., GNSS-LP links) on GNSS satellite operations.
	\end{enumerate}

	To address this gap, this paper proposes the first CPD scheme designed for joint GNSS-LP constellations.
	Specifically, the contributions are as follows:
	\begin{enumerate}
		% \item  \textbf{Topology Model}: This paper adopts the slot as the basic unit for establishing ISLs. 
		% We consider the significant orbital differences between LP satellites and GNSS satellites and introduce the concepts of \textit{LongSlots} and \textit{ShortSlots}. 
		% Earth-Moon scale links (i.e., ISLs except GNSS inner ISLs) are established based on LongSlots, whereas ISLs within the GNSS are established based on ShortSlots. 
		% Subsequently, a CPD process suitable for joint constellations is proposed. 
		% This process involves initially planning LongSlots to schedule Earth-Moon-scale links, followed by planning multiple corresponding ShortSlots during the period of these LongSlots to schedule ISLs among GNSS. 
		% As time progresses, the process of LongSlots planning and LongSlots planning is repeated. 
		% This CPD scheme addresses the link scheduling discrepancies resulting from the significant orbital differences between LP and GNSS satellites. 
		% Furthermore, the aforementioned CPD approach allows the scheduling outcomes of long and short slots to influence each other, contributing to the construction of an optimal global topology.
		\item \textbf{Hierarchical Topology Model and CPD process}:
		We introduce a novel hierarchical topology model that accounts for the orbital disparities between GNSS and LP satellites by defining two distinct link duration units: \textit{LongSlots} for Earth–Moon scale links and \textit{ShortSlots} for intra-GNSS links.
		A corresponding hierarchical and crossed CPD process is proposed, in which LongSlot-based planning governs inter-domain links, followed by nested ShortSlot planning for intra-GNSS links.
		This hierarchical and crossed scheduling allows dynamic interactions between the two layers, enabling global topology optimization while respecting the different link dynamics.
		% \item  \textbf{Joint CPD Method}: 
		% In this paper, satellites' ranging and communications potential energy and communication potential energy are formulated as ranging potential energy and communication potential energy, respectively, while users' service requirements for satellites are expressed as user potential energy. 
		% Additionally, the significant orbital differences between GNSS satellites and LP satellites/users lead to the exclusion of potential energy from GNSS satellites towards LP satellites /users. 
		% The Joint CPD ((J-CPD)) method aims to derive the ultimate topology solution by constructing a topology that maximizes joint constellation potential energy release.
		\item \textbf{Potential-Driven Joint CPD Method (J-CPD)}:  
		Adapted to the hierarchical and crossed CPD Process, we propose J-CPD, a potential-energy-inspired scheduling framework that jointly considers:  
		(\textit{i}) communication potential (based on telemetry backhaul pressure),  
		(\textit{ii}) ranging potential (for diverse orbit determination),  
		(\textit{iii}) user potential (to capture service demands), and  
		(\textit{iv}) exclusion potential (to penalize inefficient GNSS resource use).  
		By maximizing the overall release of potential energy across all these dimensions, J-CPD produces topology-aware and mission-driven link schedules.
		% \item \textbf{Performance Evaluation}: 
		% Using a simulation scenario comprising the BeiDou constellation and LP satellites deployed at L3, L4, L5, and DRO, we verified the superior performance of the proposed J-CPD. 
		% We compared the delay, ranging, and user satisfaction rate of the J-CPD method under different parameter settings with those of the FCP method~\cite{36}. 
		% The results demonstrated that J-CPD can provide excellent service while meeting the communication and ranging resources needs for GNSS and LP satellites. 
		% In contrast, the FCP approach is unsuitable for this scenario due to inefficient resource allocation. 
		% Furthermore, J-CPD exhibited varying performance under different parameter settings, indicating its adaptability as a topology construction method that can be tailored according to mission characteristics.
		\item \textbf{Comprehensive Performance Evaluation}:  
		Using a simulation based on the BeiDou GNSS and LP satellites located at L3, L4, L5, and a DRO, we evaluate J-CPD against the Fair Contact Plan (FCP) baseline~\cite{36}.  
		Results show that J-CPD significantly outperforms FCP in terms of delay reduction, ranging link diversity, while maintaining flexibility under different parameter configurations.  
		These findings confirm J-CPD's suitability for future integrated navigation and communication constellations spanning Earth–Moon space.
	\end{enumerate}

	The remainder of this paper is structured as follows:
	Section~\ref{sec_background} reviews the related work, summarizing CPD, LPs, and the interaction between LPs and GNSS. 
	Section~\ref{sec_system_model} describes the unique topology model and CPD process employed in this paper. 
	Section~\ref{sec_joint_cpd} models the diverse needs of GNSS satellites, LP satellites, and users by borrowing the concept of potential energy. It then employs maximum weight matching to identify a topology scheme that maximizes the release of constellation potential energy.
	Section~\ref{sec_evaluation} offers a performance analysis of the J-CPD proposed in this paper. 
	Finally, conclusions are drawn in Section~\ref{sec_conclusion}.
	
	\section{Background}
	\label{sec_background}
	This section provides context and related scholarly work on LPs, the interaction between LP constellation and GNSS, and Contact Plan Design. 
	
	\subsection{Libration Point}
	
	The LPs, or Lagrangian points, represent unique locations in the three-body problem. 
	At these points, the gravitational forces exerted by two large masses (e.g., the Earth and the Sun) balance with the centrifugal force experienced by a third, negligible mass (e.g., a satellite). 
	This balance enables the small object to remain stationary or maintain stable motion near these points. 
	There are five such LPs, labeled L1 through L5~\cite{1}.
	
	The development of relevant LP missions is thoroughly reviewed in~\cite{3}.
	Some typical missions include the ISEE-3 mission~\cite{4}, launched successfully on August 12, 1978, and became the first mission to enter the Sun-Earth L1 halo orbit.
	In 2020, the Chinese Chang'E-5 (CE5) service module was maneuvered into a Sun-Earth L1 LP orbit~\cite{5}.
	The utilization of lunar LPs has progressed more slowly than that of the Sun-Earth LPs.
	In 2010, a spacecraft from the Artemis~\cite{11} mission of the United States reached the Earth-Moon L2 point, marking the first triumphant arrival and exploration mission at this LP~\cite{2}.
	The Chang'E-4(CE4) relay communication satellite QueQiao is deployed at the Earth-Moon L2 point in 2018~\cite{6}.
	
	Based on the successful implementation of the aforementioned LP missions, numerous studies have investigated the deployment of deep space navigation constellations at the Earth-Moon LPs.
	The design and navigation performance of Earth-Moon LP  constellations have been examined for specific missions, such as the American Artemis program~\cite{7,8}.
	The paper in~\cite{9} analyzed the autonomous navigation performance of various Halo orbits and Distant Retrograde Orbits (DROs) combinations.
	Authors in~\cite{10} extended the navigation range to the Earth-Moon triangular LPs, placing satellites at the L4 and L5 LPs to achieve autonomous orbit determination.
	The work in~\cite{17} designs a five-satellite constellation consisting of three halo orbit satellites and two DRO satellites. 
	This constellation can provide 100\% continuous one-fold coverage of the entire lunar surface, thereby demonstrating the potential of LP-based constellations for comprehensive spatial coverage.
	
	All these studies have revealed the potential and advantages of LP constellations for future deep-space communication and navigation applications.

	\subsection{Interaction between LP constellation and GNSS}
	Much research has explored methods to combine three-body LP orbits with traditional two-body orbits (such as those used by GNSS). 
	In~\cite{12}, it is proposed that the Beidou satellites, in conjunction with the Queqiao relay satellite operating in the Earth-Moon L2 point, form an extended constellation. This configuration aims to determine a joint autonomous orbit for the Queqiao and the Beidou satellites.
	\cite{13} demonstrates that the Earth-Moon Lagrange constellation, through the establishment of inter-satellite ranging links with the BeiDou, can enable the traceability and alignment of the Earth-Moon spatiotemporal reference frame to the high-precision Earth-based spatiotemporal reference frame.
	In~\cite{14,15}, it is argued that for the LP + BeiDou satellites scenario, the augmented inter-satellite links to the BeiDou can significantly enhance the orbital accuracy of the LP constellation and reduce the required arc length for orbit determination.
	In~\cite{16}, a navigation constellation consisting of 3 GPS satellites and one LP satellite was designed. Simulation results indicate the LP satellite can achieve autonomous navigation and orbit determination using only inter-satellite ranging information. 
	The gyration drift errors of the GNSS are also mitigated.
	These studies demonstrate the numerous benefits of combining LP constellations with GNSS.

	\subsection{Contact Plan Design}
	Suppose two satellites meet certain conditions, such as no other objects obstructing the inter-satellite line of sight. 
	In that case, they are within each other's beam coverage range, and the link budget is adequate. 
	It is said that a potential link establishment opportunity (i.e., contact) exists between these two satellites. 
	Over a given period, the collection of all contacts within a constellation forms the \textit{contact topology}. 
	The process of selecting, from this contact topology, the set of links that satisfy all constraints (such as the number of terminals carried by the satellites) and will ultimately be established (the \textbf{contact plan}) is referred to as contact plan design~\cite{18}.
	
	In a navigation constellation, satellites typically carry only one ISL terminal, and each satellite must perform diverse ranging and communications within the constraints of these limited terminal resources.
	Based on heuristic algorithms:
	\cite{23} utilizes the Simulated Annealing algorithm to seek topology solutions that satisfy ranging constraints while achieving superior communication performance;
	\cite{24} proposes a high-performance memetic algorithm for generating superior topologies and a data-driven heuristic for rapid output to address the dual needs of regular scheduling and fast-response scheduling for GNSS;
	\cite{26} employs the Fireworks Algorithm to generate topologies that minimize the number of links between ground stations and satellites and inter-satellite communication costs.
	
	Based on graph theory matching:~\cite{20} uses the Non-dominated Sorting Genetic Algorithm II to generate and optimize satellite sequences for all time slots, thereby producing diverse topology solutions;
	\cite{22} adjusts the communication performance and ranging performance of GNSS constellations by modifying the application frequency of bipartite graph matching and general graph matching;
	\cite{29} achieves balanced ranging and communication performance by adjusting the edge weights between non-anchor satellites and anchor satellites;
	\cite{25} proposes the first distributed CPD scheme suitable for autonomous satellite operations through the deterministic use of matching algorithms.
	
	\cite{27} proposes a GNSS CPD scheme based on linear programming that can obtain optimal topology with minimal delay while satisfying ranging constraints. Unlike the persistent satellite-to-ground links considered in~\cite{26},~\cite{32} considers time-division satellite-to-ground links and incorporates these time-division links into the linear programming CPD process for GNSS.
	
	Strategically:
	\cite{19} introduces the concept of symbolic variance to reflect and adjust the ranging balance among satellites within the constellation.
	\cite{28} proposes a satellite layer allocation algorithm designed to achieve time synchronization across the entire constellation with the minimum number of link hops.
	\cite{30} addresses GNSS constellation communication metrics by grouping non-anchor satellites and connecting them to anchor satellites.
	
	% \subsection{The Gap in Joint CPD between GNSS and LP Constellation}
	% Despite the numerous benefits demonstrated by current research regarding the interaction between GNSS and LP constellations and the diversity of CPD schemes available for GNSS, no existing schemes are capable of simultaneous scheduling of LongSlot and ShortSlot ISLs in GNSS-LP joint constellations; nor do they account for the impact that high time cost ISL between GNSS satellites and LP satellites/users imposes on intra-GNSS resources.
	\subsection{The Gap in Joint CPD for GNSS–LP Constellations}
	While prior studies highlight the benefits of integrating GNSS and LP constellations and propose diverse CPD schemes for GNSS, none can simultaneously schedule LongSlot and ShortSlot ISLs in a GNSS–LP joint constellation.
	Moreover, existing approaches overlook the substantial time cost of GNSS–LP and GNSS–user ISLs, and the resulting impact on intra-GNSS resource availability.
	
	This paper presents the first CPD scheme that integrates LP constellations and GNSS into a single CPD process. 
	The proposed scheme is capable of meeting, balancing, and adjusting the ranging and communication needs within both GNSS and LP constellations and serving the link requirements of space users.
	
	\section{System Model}
	\label{sec_system_model}

	\subsection{Topology Model}
	\label{sec_system_model_topo}
	
	\begin{figure}[] 
		\centering
		\includegraphics[width=0.85\linewidth]{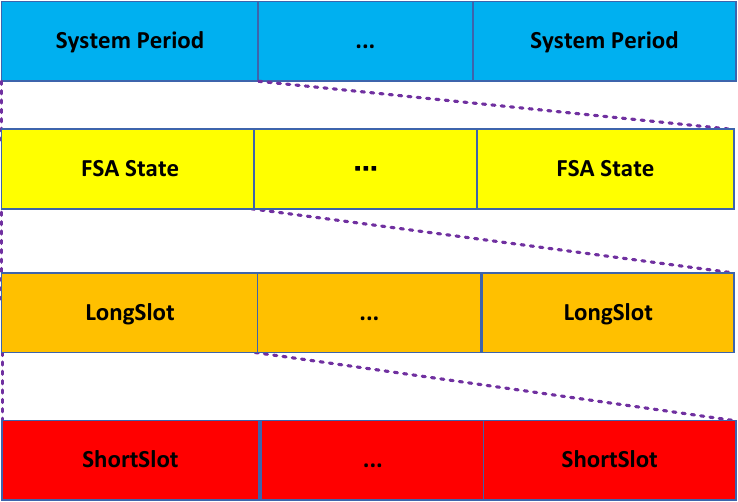}
		\caption{Hierarchical time structure of the joint constellation: System Period, FSA States, LongSlots, and ShortSlots.}
		\label{fig1}
	\end{figure}
	
	% Two satellites are considered mutually visible if there is no obstruction between them and both are within each other's antenna beam range; otherwise, they are considered invisible. 
	% Since satellites move along their trajectories, the visibility relationships among satellites change over time. 
	% We adopt a Finite State Automaton (FSA) model from~\cite{33} to address this issue. 
	% We divide the system period of the constellation into a series of equal-length states. 
	% If two satellites remain visible throughout a given state interval, they are considered visible during that state.
	To model dynamic visibility relationships in the joint constellation, we adopt a Finite State Automaton (FSA) framework~\cite{33}.
	In this model, the system period is divided into equal-length FSA states. 
	If two satellites maintain continuous mutual visibility during an FSA state—meaning unobstructed line-of-sight and mutual beam coverage—they are considered visible for that interval.
	
	% Based on FSA, we convert dynamic visibility relationships into a series of static visibility relationships. 
	% Considering that navigation satellites need to establish links with as many different objects as possible to obtain diverse ranging information, we further divide FSA states into equal-length time slots. 
	% Satellites switch their link partners at these time slots. 
	% However, given the propagation delays at Earth-Moon scales (on the order of seconds) are significantly longer than those within GNSS constellations (on the order of milliseconds), we define \textit{LongSlots} and \textit{ShortSlots}, where LongSlots are integer multiples of ShortSlots. 
	% Links at Earth-Moon scales switch in units of LongSlots, while links within the GNSS constellation switch in units of ShortSlots. 
	% Detailed information is provided in Table~\ref{Type of Links and corresponding switching unit}.
	To accommodate the differing communication timescales within the constellation, each FSA state is further divided into two types of time slots:
	\begin{itemize}
		\item \textit{LongSlots} are used for links involving LP satellites and Earth-Moon scale interactions, which involve longer propagation delays (on the order of seconds).
		\item \textit{ShortSlots} are used exclusively within the GNSS constellation, where propagation delays are significantly shorter (on the order of milliseconds).
	\end{itemize}
	
	\begin{table}[]
		\centering
		\caption{Type of Links and Corresponding Switching Unit}
		\label{Type of Links and corresponding switching unit}
		\begin{tabular}{|l|c|}
			\hline
			\textbf{Type of Links} & \textbf{Switching Unit} \\
			\hline
			ISLs between LP Satellites & LongSlot \\
			\hline
			ISLs between LP Satellites and Users & LongSlot \\
			\hline
			ISLs between GNSS Satellites and LP Satellites& LongSlot \\
			\hline
			ISLs between GNSS Satellites and Users & LongSlot \\
			\hline
			ISLs between GNSS Satellites & ShortSlot \\
			\hline
		\end{tabular}
	\end{table}
	
	% Based on these definitions, the hierarchical topology model that serves as the foundation for the remainder of this paper is depicted in Fig.~\ref{fig1}.
	This hierarchical time structure, illustrated in Fig.~\ref{fig1}, enables independent scheduling for intra-GNSS and inter-domain links between GNSS, LP satellites, and external users. 
	The link types and their corresponding switching units are summarized in Table~\ref{Type of Links and corresponding switching unit}.
	
	% Considering the rapid electrical switching capability of spaceborne phased-array antennas, link transitions at both ShortSlot and LongSlot durations can be executed with negligible time overhead. Furthermore, the inherent beam scanning range and coverage area of phased arrays antennas ensure that minor orbital deviations (on the order of hundreds of meters) of satellites or users do not significantly affect visibility relationships necessary for link establishment.
	Given the rapid electronic beam steering of spaceborne phased-array antennas, link transitions at both ShortSlot and LongSlot scales can be executed with negligible time overhead.
	Moreover, their wide beam scanning range and coverage area ensure that minor orbital deviations—on the order of hundreds of meters—do not materially affect the visibility conditions required for link establishment.
	
	\subsection{CPD Process}
	
	\begin{figure}[] 
		\centering
		\includegraphics[width=0.9\linewidth]{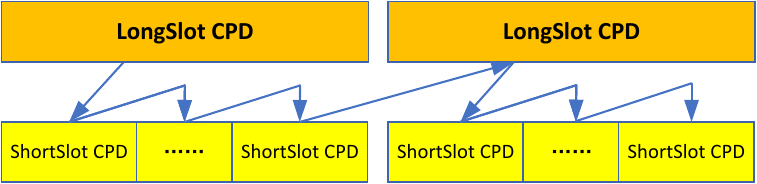}
		\caption{Hierarchical and Crossed CPD process in the joint constellation: LongSlot and ShortSlot scheduling.}
		\label{fig2}
	\end{figure}
	
	% Considering a joint constellation, different links have distinct switching units. 
	% Unlike previous studies that only conducted CPD within GNSS in ShortSlots sequentially, the CPD process adopted in this paper, as shown in Figure~\ref{fig2}, is more suitable for the joint constellation. 
	% Initially, a LongSlot CPD is performed, which plans ISLs with the switching unit being the LongSlot (i.e., Links at Earth-Moon scales). 
	% Based on the outcomes of the LongSlot CPD, sequential ShortSlot CPDs are carried out within the duration covered by the LongSlot. 
	% This involves planning ISLs with the switching unit being the ShortSlot (i.e., ISLs between GNSS Satellites). 
	% Upon completion of the ShortSlot CPDs, the timeline advances to execute the next LongSlot CPD, repeating the process.
	The joint constellation introduces heterogeneous link types, each with its own switching granularity. 
	We propose a hierarchical and crossed Contact Plan Design (CPD) process that aligns with the dual-slot structure introduced in Section~\ref{sec_system_model_topo}. 
	This approach contrasts with traditional GNSS CPD methods targeting intra-GNSS links using uniform ShortSlots.
	
	As shown in Fig.~\ref{fig2}, the proposed CPD process operates in two stages:
	\begin{enumerate}
		\item \textbf{LongSlot CPD}: At the beginning of each LongSlot interval, the CPD engine schedules all inter-domain ISLs involving LP satellites, users, and GNSS satellites—i.e., links with Earth-Moon-scale delays. These links remain fixed for the duration of the LongSlot.
		Note that each GNSS satellite is equipped with a single terminal. Thus, a GNSS satellite engaged in an ISL with an LP satellite/user during LongSlot CPD is precluded from establishing additional inter-GNSS ISLs in subsequent ShortSlot CPD.
		\item \textbf{ShortSlot CPD}: Within each LongSlot, multiple ShortSlots follow. During each ShortSlot, a separate CPD instance schedules intra-GNSS ISLs to address GNSS satellites' higher-frequency dynamics and lower-latency requirements. 
	\end{enumerate}
	
	This process iterates over the system period: each LongSlot CPD is followed by a sequence of ShortSlot CPDs, after which the next LongSlot CPD is triggered. 
	The hierarchical nature of the CPD process enables sequential scheduling of LongSlot and ShortSlot ISLs in the joint constellation. 
	Simultaneously, the crossed nature allows LongSlot and ShortSlot scheduling results to mutually constrain and iteratively optimize each other.
	For example, suppose a GNSS satellite has achieved sufficient communication and ranging performance through ShortSlot ISLs with other GNSS nodes during the ShortSlot CPD phase. 
	In that case, it becomes a stronger candidate for establishing high-time-cost LongSlot links with LP satellites/users in subsequent LongSlot CPD.
	
	\subsection{User Model}
	% In this paper, external spacecraft running in cislunar space that requires communication or navigation services via ISLs provided by GNSS satellites or LP satellites are collectively referred to as "users." 
	% Given the focus on the CPD scheme in this study, we do not differentiate the specific purposes for which satellites provide ISLs to users. 
	% Instead, we solely focus on the total number of ISLs the joint constellation provides to users.
	In this work, users refer to cislunar spacecraft requiring communication/navigation services via ISLs with GNSS/LP satellites. We treat service specifics abstractly (e.g., telemetry relay, navigation) and focus on the total ISLs delivered per FSA state by the joint constellation to quantify the CPD scheme.
	
	% Therefore, we define the format for user requirements as follows:
	% $U$=$[(x,L^u_x),\dots,(z,L^u_z)]$, where $(x,L^u_x)$ represents that user x expects satellites to provide $L^u_x$ ISLs in a FSA state for x. 
	% The joint constellation needs to provide users with as many ISLs as possible based on their individual link establishment requirements.
	% The user requirements should be set up by the ground station administrator before executing the joint CPD.
	User service requirements are formalized as:
	$U$=$[(x,L^u_x),\dots,(z,L^u_z)]$, 
	where each tuple $(x, L^u_x)$ specifies that user x expects  $L^u_x$ ISLs per FSA state. 
	These links may originate from any satellite in the GNSS or LP constellation.
	
	The J-CPD process aims to maximize successfully allocated ISLs for users subject to $U$, which is predetermined by the ground segment and remains fixed during CPD execution.

	\section{Joint CPD Model}
	\label{sec_joint_cpd}
	
	We assume that both the GNSS and the LP satellites are equipped with only one ISL terminal, and the linking parties exist in node pairs. 
	Therefore, we base our approach on graph matching methods~\cite{34}, adjusting the weights of the edges to derive an appropriate CPD scheme.
	
	We model the visibility relationships between nodes as a graph $G(V, E, W)$. 
	Here, $V$ represents the set of nodes, partitioned into subsets $V^s$, $V^l$, and $V^u$, representing GNSS satellites, LP satellites, and users, respectively. 
	$E$ denotes the set of edges representing the visible relationships between nodes. 
	$W$ is the set of weights assigned to each edge.
	
	In this section, we first describe some specific settings adopted in this paper. 
	Then, considering that this paper adopts the CPD process as shown in Fig~\ref{fig2}, we will discuss it in two parts: LongSlot CPD and ShortSlot CPD.
	
	\subsection{Potential Energy Formulations}
	\label{sec_potential_energy}
	
	To guide link scheduling decisions in the joint constellation, this paper adopts a potential energy framework inspired by classical physics. 
	Different system needs—such as telemetry data offloading, ranging diversity, and user servicing—are captured through dedicated energy models. 
	Each form of potential energy quantifies the urgency or desirability of establishing a specific link, and together they drive the edge weight assignment in the CPD optimization process.
	This section presents the mathematical formulations and parameter settings used to compute these potential energies across different link types.
	
	%\subsubsection{Communication Potential Energy}
	\subsubsection{Telemetry Offloading}

	\begin{table*}[h]
		\centering
		\caption{Communication potential energy parameters in different forms}
		\label{Communication constant values}
		\centering
		\resizebox{\linewidth}{!}
		{
			\begin{tabular}{|l|c|c|c|p{9cm}|}
				\hline
				\textbf{Type of non-anchor satellite $i$} & \textbf{Type of anchor satellite $j$} & \textbf{Parameter} & \textbf{Value}&\textbf{Explanation}\\
				\hline
				GNSS satellite & any &  $C_c^i$  & $C_c^S$ & communication constants of GNSS satellites \\
				\hline
				LP satellite & any &  $C_c^i$ & $C_c^L$&communication constants of LP satellites  \\
				\hline
				GNSS satellite & LP satellite & $h_{i,j}$ & $h_{S,L}^{n-a}$ &virtual height difference between non-anchor GNSS satellites and anchor LP satellites  \\ 
				\hline
				GNSS satellite & GNSS satellite & $h_{i,j}$  & $h_{S,S}^{n-a}$&virtual height difference between non-anchor GNSS satellites and anchor GNSS satellites\\
				\hline
				LP satellite & GNSS satellite &  $h_{i,j}$  &$h_{L,S}^{n-a}$&virtual height difference between non-anchor LP satellites and anchor GNSS satellites \\
				\hline
				LP satellite & LP satellite &  $h_{i,j}$  &$h_{L,L}^{n-a}$&virtual height difference between non-anchor LP satellites and anchor LP satellites \\
				\hline
			\end{tabular}
		}
	\end{table*}
	
	The core traffic of navigation satellites consists of frequently generated telemetry data that needs to be transmitted back to the ground stations (GSs). 
	A satellite visible to any GS is called an \textit{anchor satellite}. 
	In contrast, a satellite not visible to any GS is termed a non-anchor satellite. 
	We assume all satellites have dedicated-to-GS links, meaning anchor satellites can directly transmit data to the ground. 
	Non-anchor satellites adopt a one-hop routing strategy. 
	They can only relay telemetry data to the GS via anchor satellites and do not transmit telemetry data among themselves.
	
	%We use the concept of \textit{energy} to model the communication requirements of satellites as communication potential energy.
	We take the telemetry data offloading from satellites as the primary communication requirement. Correspondingly, we use communication potential energy to express the urgency of meeting the satellite's communication requirements.

	Let the amount of telemetry data carried by satellite $i$ at time slot $k$  be denoted as $D(i,k)$. 
	We assume that each LP/ GNSS satellite generates one unit of telemetry data at the beginning of every LongSlot/ ShortSlot, so:  
	\begin{equation}
		D(i,k+1)=D(i,k)+1, \label{eq0}
	\end{equation}
	
	Since anchor satellites can directly transmit data to the ground, their $D(i,k)$ is always set to 0.
	
	For non-anchor satellites, before establishing an ISL with an anchor satellite, the amount of telemetry data $D(i,k)$ will accumulate over time slots. 
	When such a link is formed in slot $k$, the non-anchor satellite transfers all $D(i,k)$ to the anchor for ground transmission, resetting $D(i,k)$ to zero by the end of the slot.
	
	Drawing inspiration from \textit{gravitational potential energy $E=mgh$}, we define the communication potential energy of a non-anchor satellite $i$ relative to an anchor satellite $j$ at time slot $k$ as:
	\begin{equation}
		E_c^{i,j,k} = C_c^i*D(i,k)*h_{i,j}, \label{eq1}
	\end{equation}
	where $C_c^i$ is a communication constant analogous to the gravitational constant $g$. 
	$D(i,k)$ represents the accumulated telemetry data of non-anchor satellite $i$ at time slot $k$, analogous to mass $m$ in gravitational potential energy. 
	$h_{i,j}$ denotes the virtual height difference between non-anchor satellite $i$ and anchor satellite $j$, analogous to the height $h$ in gravitational potential energy.
	
	Additionally, the values of the $C_c^i$ and $h_{i,j}$ in Equation~\eqref{eq1} vary depending on the types of non-anchor satellite $i$ and anchor satellite $j$. 
	The specific values are detailed in Table~\ref{Communication constant values}.
	The virtual height difference governs satellite preferences for downlinking telemetry data. For instance, when $h_{L,S}^{n-a}>h_{L,L}^{n-a}$ as specified in Table~\ref{Communication constant values}, non-anchor LP satellites preferentially relay data through anchor GNSS satellites over anchor LP satellites.
	
	It should be noted that communication potential energy exists only between non-anchor satellites relative to anchor satellites. 
	The communication potential energy between non-anchor satellites, among anchor satellites, and from an anchor satellite relative to a non-anchor satellite is considered zero.
	
	%\subsubsection{Ranging Potential Energy}
	\subsubsection{Ranging Diversity}
	
	\begin{table*}[h]
		\centering
		\caption{Ranging potential energy parameters in different forms}
		\label{ranging constant values}
		\begin{tabular}{|l|c|c|c|c|p{8cm}|}
			\hline
			\textbf{Type of satellite i} & \textbf{Type of satellite j} & \textbf{Parameter} & \textbf{Value}&\textbf{unit} &  \textbf{Explanation} \\
			\hline
			GNSS satellite & any & $C_r^i$   & $C_r^S$ & / & The ranging constants of GNSS satellites \\
			\hline
			LP satellite & any & $C_r^i$ & $C_r^L$ & / &The ranging constants of LP satellite\\
			\hline
			GNSS satellite & any & $N_i$ & $N_S$ & 1 & The desired number of ranging links for GNSS satellites\\
			\hline
			LP satellite & any & $N_i$ & $N_L$ & 1 &The desired number of ranging links for LP satellites\\
			\hline
			LP satellite & any & $n~and~m$  & / & LongSlot & time slot index related to LP satellites are conducted using LongSlot.\\
			\hline
			GNSS satellite & GNSS satellite &  $n~and~m$  & / & ShortSlot & time slot index between GNSS satellites are conducted using ShortSlot\\
			\hline
			LP satellite & any &  $B_i$  &$B_r^L$ & / &The ranging bias constants of LP satellites \\
			\hline
			GNSS satellite & any &  $B_i$  &$B_r^S$ & / &The ranging bias constants of GNSS satellites \\
			\hline
			LP satellite & any &$I_i$ & $I_L$ &LongSlot& The interval number of LongSlots for effective ranging at the Earth-Moon scale\\
			\hline
			GNSS satellite & GNSS satellite & $I_i$ & $I_S$ &ShortSlot& The interval number of ShortSlots for effective ranging at the GNSS scale \\
			\hline
		\end{tabular}
	\end{table*}
	
	\begin{table}[h]
		\centering
		\caption{Initial ranging time slot index set}
		\label{initial ranging time slot index set}
		\begin{tabular}{|l|c|c|}
			\hline
			\textbf{Type of satellite i} & \textbf{Type of satellite j} & \textbf{initial ranging index $m_{init}$} \\
			\hline
			GNSS satellite & GNSS satellite & -$I_S$ ShortSlots\\
			\hline
			LP satellite & LP satellite & -$I_L$ LongSlots \\
			\hline
			GNSS satellite & LP satellite & -$I_L$ LongSlots \\
			\hline
		\end{tabular}
	\end{table}
	For orbit determination and time synchronization, satellites must establish links with as many other satellites as possible to obtain diverse-ranging data. 
	In previous research on GNSS CPD, a ranging link between a satellite and other satellites within a specific period was considered adequate if it was unique, i.e., non-repetitive~\cite{19,27,29}. 
	
	This paper defines an effective ranging link differently: If the time elapsed since the last ranging link between satellite $i$ and satellite $j$ exceeds a specific interval, the newly established link between these two satellites is considered an effective ranging link. 
	The essence of this definition aligns with the previous notion of non-repetitive links, aiming to ensure minimal correlation among ranging data.
	
	Assuming that the last established ranging link between satellite $i$ and $j$ was at time slot $m$. 
	The current CPD process is at time slot $n$, and the ranging potential energy between satellite $i$ and $j$ is defined analogously to \textit{elastic potential energy $E=\frac{1}{2}kx^2$}. 
	
	Specifically, the ranging potential energy of satellite $i$ relative to satellite $j$ at slot n is defined as:
	\begin{equation*}
		E_r^{i,j,n}=\left\{ \begin{matrix} C_r^i(N_i^{~}-N_i^{got})(n-m-I_i)+B_i, &  condition1\\ 0, & others &\\ \end{matrix} \right. 
	\end{equation*}
	\begin{equation}
		~~~~~~~~~condition1: n-m > I_i~and~N_i^{got} < N_i, \label{eq2}
	\end{equation}
	where $C_r^i$ denotes the ranging constant. 
	$N_i$ represents the number of effective ranging links that satellite $i$ aims to establish within an FSA state.
	$N_i^{got}$ indicates the number of effective ranging links that satellite $i$ has already established up to the current time slot $n$.
	$(n-m)$ signifies the interval, in terms of time slots, between the current time slot $n$ and the last time slot $m$ when satellites $i$ and $j$ successfully establish a ranging link.
	$B_i$ is the ranging bias.
	$condition1$ states that more than $I_i$ time slots have passed since the last ranging link establishment between satellites $i$ and $j$, and the number of acquired ranging links $N_i^{got}$ for satellite $i$  is less than its desired number of ranging links $N_i$. In other words, $I_i$ defines the time interval required for effective ranging links.
	
	When $condition1$ is satisfied, it is considered that satellite $i$ tends to establish a ranging link with satellite $j$, indicating the presence of ranging potential energy. 
	Analogous to the elastic potential energy in a spring, where $x$ represents displacement, the larger the value of $x$, the more stretched the spring becomes, thus the more significant the elastic potential energy. 
	Similarly, the larger the value of $(n-m)$, the lower the correlation between the data obtained from this range and the previous one, leading to a higher potential energy.
	The term $(N_i^{~}-N_i^{got})$ can be compared to the spring constant $k$ in Hooke's law. In reality, the more frequently a spring is stretched, the looser it becomes over time, decreasing its spring constant. Likewise, as $N_i^{got}$ increases, the number of established ranging links by satellite $i$ approaches its expected value $N_i$, causing $(N_i^{~}-N_i^{got})$ to gradually decrease, which in turn reduces the ranging potential energy.
	
	Similarly to communication potential energy, the values of the constants in Equation~\eqref{eq2} vary depending on the types of satellites i and j. 
	The specific values are detailed in Table~\ref{ranging constant values}.
	Unlike communication potential, ranging potential exists between any satellite and any other satellite.
	
	Additionally, we assume the initial ranging time slot  $m_{init}$  between satellites as shown in Table~\ref{initial ranging time slot index set}. 
	This setup allows each satellite to establish an effective ranging link from the first time slot.
	
	%\subsubsection{User Potential Energy}
	\subsubsection{User Service Demand}
	
	This paper primarily focuses on the number of links the satellite constellation provides to users. 
	Considering these links may be used for navigation services (fundamentally based on ranging), we still simulate elastic potential energy to define user potential energy, ensuring high-quality ranging from satellites to users.
	Specifically, the user potential energy of user $i$ relative to satellite $j$ at slot n is defined as:
	\begin{equation*}
		E_u^{i,j,n}=\left\{ \begin{matrix} C_U(L_u^i-N_i^{got})(n-m-I_U)+B_U, &  condition1\\ 0, & others &\\ \end{matrix} \right. 
	\end{equation*}
	\begin{equation}
		~~~~~~~~~condition1: n-m > I_U~and~N_i^{got} < L_u^i,  \label{eq3}
	\end{equation}
	where $C_U$ is a user service constant, and $B_U$ denotes the service bias for the user. 
	$n$ represents the current LongSlot index, while $m$ indicates the LongSlot index when satellite $j$ last provided a link to user $i$. 
	$I_U$ is the number of LongSlots required for effective links. $L_u^i$ stands for the number of link establishments requested by user request $U$, whereas $N_i^{got}$ denotes the total number of links established by user $i$ with all satellites up to the current LongSlot $n$.
	
	As the number of links the satellite provides increases, the user potential decreases.  
	Users exhibit higher user potential values for satellites with which they have not established a link for a long time.
	
	Additionally, assuming the initial link establishment time slot $m_{init}$ between users and each satellite is  -$I_U$ LongSlots, this configuration allows satellites to serve users from the first time slot.

	%\subsubsection{GNSS Exclusion Potential Energy}
	\subsubsection{GNSS Internal Resource Protection}
	
	\begin{table*}[h]
		\centering
		\caption{Exclusion potential energy parameters in different forms}
		\label{exclusion constant values}
		\begin{tabular}{|l|c|c|c|p{8cm}|}
			\hline
			\textbf{GNSS satellite i} & \textbf{Type of external object j} & \textbf{Parameter} & \textbf{Value} &  \textbf{Explanation} \\
			\hline
			any & LP satellites & $C_e^j$   & $C_e^L$  & Exclusion constant of GNSS satellites relative to satellite at LPs \\
			\hline
			any & Users & $C_e^j$ & $C_e^U$  &Exclusion constant of GNSS satellites relative to users\\
			\hline
			any & LP satellites & $B_e^j$ & $B_e^L$  & Exclusion bias of GNSS satellites relative to satellites at LPs\\
			\hline
			any & Users & $B_e^j$ & $B_e^U$  & Exclusion bias of GNSS satellites relative to users\\
			\hline
			
		\end{tabular}
	\end{table*}
	
	Inter-domain ranging/communication links between GNSS and LP satellites operate within LongSlots, while intra-GNSS links utilize ShortSlots (subunits of a LongSlot). 
	Establishing links between GNSS and LP satellites incurs significant time costs. 
	Similarly, when GNSS satellites provide services to users in units of LongSlots, the link with the user cannot perform ranging or communications for the GNSS satellite, resulting in substantial costs for the GNSS satellite to serve its users. 
	Thus, we define the exclusion potential energy for each GNSS satellite to balance resource contention caused by LongSlot ISLs: a higher value penalizes its selection for high-cost LongSlot ISLs with LP satellites/users.
	
	For a non-anchor satellite $i$, the Exclusion Potential Energy it exerts on a LP satellite or a user $j$ is at slot n defined as follows:
	\begin{equation*}
		E_e^{i,j,n}=\left\{ \begin{matrix} C_e^j(D(i,n)+N_i-N_i^{got})+B_e^j, &  condition1\\ 0, & others &\\ \end{matrix} \right. 
	\end{equation*}
	\begin{equation}
		~~~~~~~~~condition1: C_e^j(D(i,n)+N_i-N_i^{got})+B_e^j>0,  \label{eq4}
	\end{equation}
	where $C_e^j$ denotes the exclusion constant of a GNSS satellite concerning an external object $j$. 
	$D(i,n)$ represents the amount of telemetry data accumulated by non-anchor GNSS satellite $i$ till the current time slot $n$.
	$N_i$ indicates the number of ranging links that GNSS satellite $i$ is expected to establish, whereas $N_i^{got}$ denotes the number of ranging links that have been completed by GNSS satellite $i$ up to time slot $n$.
	$B_e^j$ represents the exclusion bias of a GNSS satellite concerning the external object $j$.
	
	Similarly, the value of $C_e^j$ and $B_e^j$ parameters varies depending on the external object type $j$, as detailed in Table~\ref{exclusion constant values}. 
	The value of $N_i$, according to Table~\ref{ranging constant values}, is taken as $N_S$.
	
	For non-anchor GNSS satellites, a greater accumulated telemetry data volume and a more significant number of incomplete ranging links result in a higher exclusion potential energy exerted outward.
	
	For GNSS anchor satellite $i$, since its carried telemetry data volume $D(i,n)$ is always zero, the exclusion potential energy of GNSS anchor satellite $i$ for the LP satellite or user $j$ at slot n is defined as:
	\begin{equation*}
		E_e^{i,j,n}=\left\{ \begin{matrix} C_e^j(\overline{D(i,n)}+N_i-N_i^{got})+B_e^j, &  condition1\\ 0, & others &\\ \end{matrix} \right. 
	\end{equation*}
	\begin{equation}
		~~~~~~~~~condition1: C_e^j*(\overline{D(i,n)}+N_i-N_i^{got})+B_e^j>0,  \label{eq5}
	\end{equation}
	where $\overline{D(i,n)}$ represents the average telemetry data volume carried by all other GNSS non-anchor satellites that are visible to GNSS anchor satellite $i$. The definitions of all other parameters in~\eqref{eq5} are the same as in~\eqref{eq4}.
	
	For GNSS anchor satellite $i$, its exclusion potential energy depends not only on its number of incomplete ranging links but also on the communication status with the non-anchor satellites that are visible to it.
	
	%Based on the aforementioned energy settings, this paper conducts edge weight assignments during the LongSlot CPD and ShortSlot CPD processes, thereby determining the final topology solution.
	Together, these potential functions—telemetry offloading, ranging diversity, user service demand, and GNSS internal resource protection—jointly define the edge weights in the CPD graph. They guide the link scheduling process during both LongSlot and ShortSlot phases, enabling the construction of dynamic topologies that balance internal coordination with external service delivery.
	
	\subsection{LongSlot CPD}
	
	In the LongSlot CPD process, only links at the Earth-Moon scale are considered, without considering links within the GNSS constellation. 
	Thus, we exclude the visibility relationships among GNSS satellites from graph $G(V,E,W)$ to obtain $G_{Long}(V, E, W)$. In $G_{Long}(V, E, W)$, 
	$V$ remains unchanged with graph $G(V,E,W)$. 
	$E$ denotes the set of edges representing the visible relationships between LP satellites, LP satellites and users, GNSS satellites and LP satellites, and GNSS satellites and users. $W$ is the set of weights assigned to each edge.
	
	\subsubsection{Weights Assignment for Edges}
	We define the edge weight between GNSS satellite $i$ and user $j$ at LongSlot $n$ in $G_{Long}(V, E, W)$ as follows:
	\begin{equation}
		w_{i,j,n}^{s-u}=E_u^{j,i,n}-E_e^{i,j,n} \label{eq6}
	\end{equation}
	$E_u^{j,i,n}$ represents the user potential energy of user $j$ relative to GNSS satellite $i$, and $E_e^{i,j,n}$ denotes the exclusion potential energy of GNSS satellite $i$ relative to user $j$. The specific definitions are provided in Eq~\eqref{eq3}\eqref{eq4}\eqref{eq5} and Table ~\ref{exclusion constant values}.
	
	Following this, we define the edge weight between the LP satellite $i$ and user $j$ at LongSlot $n$ as being constituted solely by the user's potential energy:
	\begin{equation}
		w_{i,j,n}^{l-u}=E_u^{j,i,n} \label{eq7}
	\end{equation}
	The edge weight between LP satellites at LongSlot $n$ is defined by the communication potential energy and the ranging potential energy:
	\begin{equation}
		w_{i,j,n}^{l-l}=E_c^{i,j,n}+E_c^{j,i,n}+ E_r^{i,j,n}+E_r^{j,i,n} \label{eq8}
	\end{equation}
	
	The edge weight between an LP satellite $i$ and a GNSS satellite $j$ is defined by three components: the communication potential energy, the ranging potential energy, and the exclusion potential energy:
	\begin{equation}
		w_{i,j,n}^{l-s}=E_c^{i,j,n}+E_c^{j,i,n}+E_r^{i,j,n}+E_r^{j,i,n}-E_e^{i,j,n} \label{eq9}
	\end{equation}
	
	There are no edges between GNSS satellites in $G_{Long}(V, E, W)$. Therefore, it is unnecessary to allocate edge weights between GNSS satellites.
	
	\subsubsection{Execute Maximum Weight Matching}
	At LongSlot $n$, after assigning the weights to all edges in $G_{Long}(V, E, W)$ according to \eqref{eq6}\eqref{eq7}\eqref{eq8}\eqref{eq9},  execute the maximum weight matching algorithm~\cite{35}. The maximum weight matching algorithm finds a matching in the weighted graph that maximizes the sum of the weights of the selected edges.
	High potential energy is inherently unstable. In this paper, the maximum weight matching naturally means maximizing the release of potential energy within the joint constellation.
	Assume that the matching result obtained in LongSlot $n$ is denoted as $M_n^{Long}$. Based on this matching result, we can directly derive the link scheduling results at the Earth-Moon scale at the LongSlot $n$.
	
	\subsubsection{PostProcess}
	After obtaining the matching result for LongSlot $n$, we need to update the potential energies between nodes according to the established links.
	Specifically, this includes:
	\begin{enumerate}
		\item \textbf{Resetting Telemetry Data for Non-anchor Satellites}: If there exists a match between a non-anchor satellite $i$ and an anchor satellite, set the telemetry data $D(i,n)$ of the non-anchor satellite $i$ to zero.
		\item \textbf{Updating Ranging Records for Satellites:}: If the link between satellite $i$ and satellite $j$ is an effective ranging link, update the ranging counts $N_i^{got}$=$N_i^{got}+1$ and $N_j^{got}$=$N_j^{got}+1$. Record the current time slot n as the latest ranging record for satellite $i$ and $j$.
		\item \textbf{Updating Service Records for User}: If there exits a link between satellite $i$ and user $j$, update the service count $N_j^{got}$=$N_j^{got}+1$ for user $j$. Record the current time slot $n$ as the latest effective service record for satellite $i$ providing links to user $j$.
		\item \textbf{Recording GNSS Satellite Information}: Record the GNSS satellites that have established links with users or LP satellites. Since each GNSS satellite carries only one terminal, during the corresponding ShortSlot CPD process, these GNSS satellites are considered invisible to all other GNSS satellites to prevent establishing links with other GNSS satellites.
	\end{enumerate}
	
	\subsection{ShortSlot CPD}
	During the ShortSlot CPD process, only the links within the GNSS constellation are considered without considering the links at the Earth-Moon scale. 
	The visibility relationships within the GNSS constellation are modeled as a graph $G_{Short}(V,E,W)$. 
	Additionally, all visibility relationships of GNSS satellites that establish links with LP satellites and users during the LongSlot CPD process also need to be excluded in $G_{Short}(V,E,W)$.
	
	\subsubsection{Weights Assignment for Edges}
	In ShortSlot $k$, the weight of an edge between satellites within the GNSS constellation is defined as:
	\begin{equation}
		w_{i,j,k}^{s-s}=E_c^{i,j,k}+E_c^{j,i,k}+E_r^{i,j,k}+E_r^{j,i,k}\label{eq10}
	\end{equation}
	Where: $E_c^{i,j,k}$ and $E_c^{j,i,k}$ denote the communication potential energy between GNSS satellites, $E_r^{i,j,k}$ and $E_r^{j,i,k}$ denote the ranging potential between GNSS satellites. The specific values are given by Equation~\eqref{eq1}\eqref{eq2} and Table~\ref{Communication constant values} and~\ref{ranging constant values} 
	
	\subsubsection{Execute Maximum Weight Matching}
	Under ShortSlot $k$,  after assigning the edge weights in a graph $G_{Short}(V,E,W)$ according to Equation~\eqref{eq10}, the maximum weight matching is performed. 
	Assuming the CPD result under ShortSlot $k$ is $M_k^{Short}$. Based on this matching result, we can obtain the link scheduling outcomes within the GNSS constellation for ShortSlot $k$.
	
	\subsubsection{PostProcess}
	After obtaining the matching result for ShortSlot $k$, we need to update the potential energies between nodes according to the established links.
	Specifically, this includes:
	\begin{enumerate}
		\item \textbf{Resetting Telemetry Data for Non-anchor GNSS Satellites}: If there exists a match between a non-anchor GNSS satellite $i$ and an anchor GNSS satellite, set the telemetry data $D(i,k)$ of the non-anchor GNSS satellite $i$ to zero.
		\item \textbf{Updating Ranging Records for GNSS Satellites:}If the link between GNSS satellite $i$ and GNSS satellite $j$ is an effective ranging link, update the ranging counts $N_i^{got}$=$N_i^{got}+1$ and $N_j^{got}$=$N_j^{got}+1$. Record the current time slot k as the latest ranging record for satellite $i$ and $j$.
	\end{enumerate}
	
	\subsection{J-CPD Algorithm}
	
	For a given FSA state comprising $N$ LongSlots, each LongSlot contains $M$ ShortSlots.
	Algorithm~\ref{alg1} illustrates the J-CPD process based on LongSlot CPD and ShortSlot CPD.
	Clearly, the J-CPD scheme fundamentally integrates a hierarchical and crossed CPD process with energy-driven heuristic weighting.
	
	\begin{algorithm} 
		\caption{J-CPD Process}
		\label{alg1}
		\begin{algorithmic}[1]
			\Require GNSS constellation parameters, LP constellation parameters, Users orbit, User requirement $U$,  Ground station locations, The values of parameters in all potential energy formula.
			\Ensure Topology solution 
			\State \textbf{Begin}
			\State $n$=1, $k=M\cdot n-(M-1)$
			\State Get graph $G_{Long}(V,E,W)$ at LongSlot $n$
			\While {$n \leq N$ }
			\State Carry out \textbf{LongSlot CPD} at LongSlot $n$
			\State Get matching result $M_n^{Long}$
			\State Set $i=1$
			\While {$i \leq M$}
			\State Get graph $G_{Short}(V,E,W)$ at ShortSlot $k$
			\State Carry out \textbf{ShortSlot CPD} at ShortSlot $k$
			\State Get matching result $M_k^{Short}$
			\State $k=k+1$, $i=i+1$
			\EndWhile
			\State $n=n+1$
			\EndWhile 
			\State Construct topology solution from All $M_n^{Long} , where~1\leq n\leq N$ and $M_k^{Short} , where ~1\leq k\leq N \cdot M$
			\State \textbf{End}
			
		\end{algorithmic}
	\end{algorithm}
	
	\section{Evaluation}
	\label{sec_evaluation}
	% This section evaluates the performance of the proposed Joint CPD method across various simulation scenarios.
	% This paper introduces novel joint LP-GNSS constellations for which no direct comparison algorithms currently exist. 
	% To demonstrate the superiority of the Joint CPD algorithm in terms of edge weight allocation, we still adopt the CPD process illustrated in Fig~\ref{fig2}, but employing the Fair Contact Plans (FCP) algorithm~\cite{36} in both the LongSlot CPD and ShortSlot CPD process to achieve fair link establishment.
	% Additionally, to adapt the FCP algorithm to the present scenario, we have modified it so that once satellites have fulfilled a user's link requirements, the FCP will cease to provide further connections for that user.
	This section evaluates the proposed J-CPD scheme's performance under various simulation scenarios.
	
	\paragraph{Evaluation Baseline}
	As this work introduces a novel joint architecture integrating GNSS and LP satellites, no existing CPD algorithms are directly applicable for baseline comparison. 
	Nevertheless, to assess the effectiveness of our edge weight allocation strategy, we adopt the same two-level CPD process illustrated in Fig.\ref{fig2}, using the Fair Contact Plan (FCP) algorithm\cite{36} as a benchmark in both LongSlot and ShortSlot scheduling phases.
	The FCP algorithm is adapted to the joint constellation scenario to ensure a fair comparison. 
	Specifically, once a user’s requested number of ISLs has been satisfied, FCP is modified to cease further link assignments to that user, thereby preserving link resources for inter-satellite operations.
	
	\paragraph{Scenario Parameters}
	In our simulation, we utilize the BeiDou system as the GNSS constellation, consisting of 24 MEO satellites, 3 GEO satellites, and 3 IGSO satellites. 
	For the LP constellation, we adopt the configuration from Reference~\cite{14}, placing one satellite each at the Earth-Moon L3, L4, and L5 points, as well as one satellite in Distant Retrograde Orbit (DRO) around the Moon. 
	The ground stations for the joint constellation are deployed in Kashgar, Jiamusi, and Sanya. 
	Additionally, $|U|$ represents the number of users in $U$, and we study scenarios with 48, 56, 64, and 72 users. 
	For this simulation, it is set by default that all users request four links (i.e., $\forall (x,L_x^u) \in U$, $L_x^u$=4) and are evenly distributed at L3, L4, L5 point, and DRO (thus we study 12, 14, 16, and 18 users at each position).
	Detailed orbital information for the constellations and coordinates of the ground stations are provided in Table~\ref{Orbit parameters and GS locations}.
	
	\begin{table}[]
		\caption{Scenario Parameters}
		\label{Orbit parameters and GS locations}
		\centering
		\footnotesize
		\renewcommand{\arraystretch}{1.2}
		\begin{tabular}{|c|c|}
			\hline
			\textbf{Satellites and GSs} & \textbf{Description} \\
			\hline
			MEO & Walker-$\delta$ 24/3/1, h = 21528 km, inc = 55° \\ 
			\hline
			IGSO & h = 35786 km, inc = 55°, interval = 120° \\
			\hline
			GEO & h = 35786 km, lon = (80°, 110.5°, 140°) \\
			\hline
			LP & L3, L4, L5, DRO \\
			\hline
			Jiamusi & (46.8°N, 130.3°E) \\
			\hline
			Kashi & (39.47°N, 75.99°E) \\
			\hline
			Sanya & (18.23°N, 109.02°E) \\
			\hline
			User Count & $|U| \in \{48, 56, 64, 72\}$ \\
			\hline
			User Orbit & L3, L4, L5, and DRO (4 links) \\
			\hline
		\end{tabular}
	\end{table}
	
	\paragraph{Basic CPD Parameters}
	The remaining basic simulation parameters are summarized in Table~\ref{Basic parameters in the simulation}. 
	Setting the ShortSlot duration for intra-GNSS constellation links to 3 seconds is a conventional practice~\cite{22,25,27,37}. 
	The LongSlot duration is set to 9 seconds, which is an integer multiple of the ShortSlot and ensures sufficient time for bidirectional pseudo range ranging, even considering propagation delays of 1 to 3 seconds at the Earth-Moon scale. 
	An FSA state is configured to last 6 minutes, encompassing 40 LongSlots and 120 ShortSlots.
	
	\begin{table}[]
		\caption{Basic CPD Parameters}
		\label{Basic parameters in the simulation}
		\centering
		\centering
		\footnotesize % 调整字体大小以适应表格宽度
		\renewcommand{\arraystretch}{1.2} % 可选：增加行高以提高可读性
		\begin{tabular}{|c|c|}
			\hline
			Simulation Duration &  7 days\\
			\hline
			Length of an FSA state & 6 min   \\ 
			\hline
			Length of a LongSlot & 9s  \\
			\hline
			Length of a ShortSlot & 3s   \\
			\hline
			ISL pointing range in MEO & 60° \\
			\hline
			ISL pointing range in GEO/IGSO &  45°\\
			\hline
			ISL pointing range in LP & 75° \\
			\hline
			GS pointing range&  85°\\
			\hline
			$I_U$, $I_L$ &  20  LongSlots\\
			\hline
			$I_S$ &  19 ShortSlots\\
			\hline
			$N_S$, $N_L$ &  60, 5 \\
			\hline
		\end{tabular}
	\end{table}
	
	The value of $I_S$ is set to 19, indicating that an interval exceeding 19 ShortSlots for inter-GNSS satellite links is considered an effective ranging link. 
	This aligns with previous studies~\cite{27,37} where a planning unit of 1 minute (i.e., 20 ShortSlots) was used, and a non-repeating link within this unit was deemed effective ranging. 
	The value of $I_L$ is set to 20 LongSlots, meaning that for links between LP satellites and between GNSS and LP satellites, an interval exceeding 20 LongSlots (equivalent to 60 ShortSlots) is required for the link to be considered effective ranging. 
	This is due to the slower changes in geometric configurations among satellites at the Earth-Moon scale, necessitating longer intervals to reduce correlations among ranging data.
	$I_U$ is also set to 20 LongSlots, indicating that the same satellite needs to wait for at least 20 LongSlots before providing another effective link to users. 
	This enhances users' navigation service performance.
	
	$N_S$, the number of ranging links required by a GNSS satellite within one FSA state, has been traditionally assumed to be 10 links per 1-minute planning unit in previous studies~\cite{27,37}. 
	Proportionally, within a 6-minute FSA state, a GNSS satellite would require 60 ranging links. 
	Reference~\cite{14} suggests that four LP satellites should establish links with each other once every 10 minutes (i.e., three ranging links per 10 minutes). 
	More strictly, this paper sets the number of ranging links required by LP satellites within one FSA state (6min), $I_L$, to 5.
	
	\paragraph{Potential Energy Parameters}
	The J-CPD scheme leverages an energy-inspired abstraction to model communication, ranging, service demands and GNSS resource contention as different forms of potential energy (communication, ranging, user, and exclusion potential energies). 
	As discussed in Section~\ref{sec_potential_energy}, each type of potential energy depends on tunable parameters that influence the relative priority of different link types during scheduling.
	
	Based on equation~\eqref{eq0}-\eqref{eq5}, we propose three parameter tuning principles:
	\begin{enumerate}
		\item \textbf{Balanced Ranging and Communication Potential}: Maintain comparable magnitude between ranging potential $E_r$ and communication potential $E_c$ for each satellite. 
		This balance enables alternating prioritization of ranging versus communication demands during scheduling.
		
		\item \textbf{User Potential Scaling}: Set user potential $E_u$ slightly lower than satellitete ranging potential $E_r$ and communication potential $E_c$.
		This helps prioritize satellite-centric objectives (ensuring robust internal ranging and communication performance) while still effectively serving users.
		\item \textbf{Adjustable GNSS Exclusion Potential}: Tune GNSS exclusion potential $E_e$ based on system goals: Excessively high $E_e$ prioritizes intra-GNSS operations, inhibiting inter-domain LongSlot ISLs; $E_e$ moderately exceeding typical $E_c/E_r$ optimally reserves GNSS resources for core needs before allocating spare capacity to LP satellite/user links.
	\end{enumerate}
	
	Following these principles, Table~\ref{potential parameters specific values} presents three representative parameter groups employed in our simulations.
	Three parameter groups (Group 1, Group 2, and Group 3) are defined to explore how changes in potential energy weights impact the system’s behavior and performance.
	These parameter groups are varied in the evaluation to test the flexibility and adaptability of the J-CPD scheme across different system priorities (e.g., favoring GNSS operations vs. user service vs. LP cooperation).
	Specifically, the second group of parameters adopted by J-CPD increases the exclusion potential ($C_e^L$ and $B_e^L$) from GNSS satellites to LP satellites compared to the first group. 
	The third group further enhances the user potential ($C_U$) relative to the second group.

	\begin{table}[t]
		\caption{Potential Energy Parameters.}
		\label{potential parameters specific values}
		\centering
		\footnotesize
		\renewcommand{\arraystretch}{1.2}
		\begin{tabular}{|c|c|c|c|}
			\hline
			\textbf{Parameter} & \textbf{Group 1} & \textbf{Group 2} & \textbf{Group 3} \\ 
			\hline
			\multicolumn{4}{|c|}{{Communication constants (GNSS/LP)}} \\
			\hline
			$C_c^S$ & 200 & 200 & 200 \\ \hline
			$C_c^L$ & 100 & 100 & 100 \\
			\hline
			\multicolumn{4}{|c|}{{Virtual height differences}} \\
			\hline
			$h_{S,L}^{n-a}$ & 2 & 2 & 2 \\ \hline
			$h_{S,S}^{n-a}$ & 7 & 7 & 7 \\ \hline
			$h_{L,S}^{n-a}$ & 5 & 5 & 5 \\ \hline
			$h_{L,L}^{n-a}$ & 5 & 5 & 5 \\
			\hline
			\multicolumn{4}{|c|}{{Ranging constants and biases}} \\
			\hline
			$C_r^S$ & 15 & 15 & 15 \\ \hline
			$C_r^L$ & 20 & 20 & 20 \\ \hline
			$B_r^S$ & 800 & 800 & 800 \\ \hline
			$B_r^L$ & 500 & 500 & 500 \\
			\hline
			\multicolumn{4}{|c|}{{Exclusion constants and biases (GNSS $\rightarrow$ LP/users)}} \\
			\hline
			$C_e^L$ & 120 & 170 & 170 \\ \hline
			$C_e^U$ & 150 & 150 & 150 \\ \hline
			$B_e^L$ & 200 & 500 & 500 \\ \hline
			$B_e^U$ & 100 & 100 & 100 \\
			\hline
			\multicolumn{4}{|c|}{{User service constants and bias}} \\
			\hline
			$C_U$ & 10 & 10 & 100 \\ \hline
			$B_U$ & 300 & 300 & 300 \\
			\hline
		\end{tabular}
	\end{table}
	
	\subsection{J-CPD Performance Evaluation}
	In this section, we compare the performance of J-CPD under three groups of parameters in Table~\ref{potential parameters specific values} and FCP with varying numbers of users. 
	To comprehensively assess the J-CPD algorithm, we analyze its performance regarding communication delay, ranging capability, user satisfaction, structural composition of established links and the computational cost of J-CPD.
	
	\subsubsection{Delay}
	In the scenario discussed in this paper, the slot duration is greater than the propagation and transmission delays of the signals. 
	Therefore, we define the delay as the number of slots that a non-anchor satellite needs to wait before linking to an anchor satellite. 
	It is noted that if a non-anchor GNSS satellite links to an LP anchor satellite, it is assumed that the non-anchor GNSS satellite can transmit data back to ground stations throughout the entire LongSlot duration.
	Since anchor satellites have dedicated space-to-ground links, their delay is 0.
	
	\begin{figure}[] 
		\centering
		\includegraphics[width=0.9\linewidth]{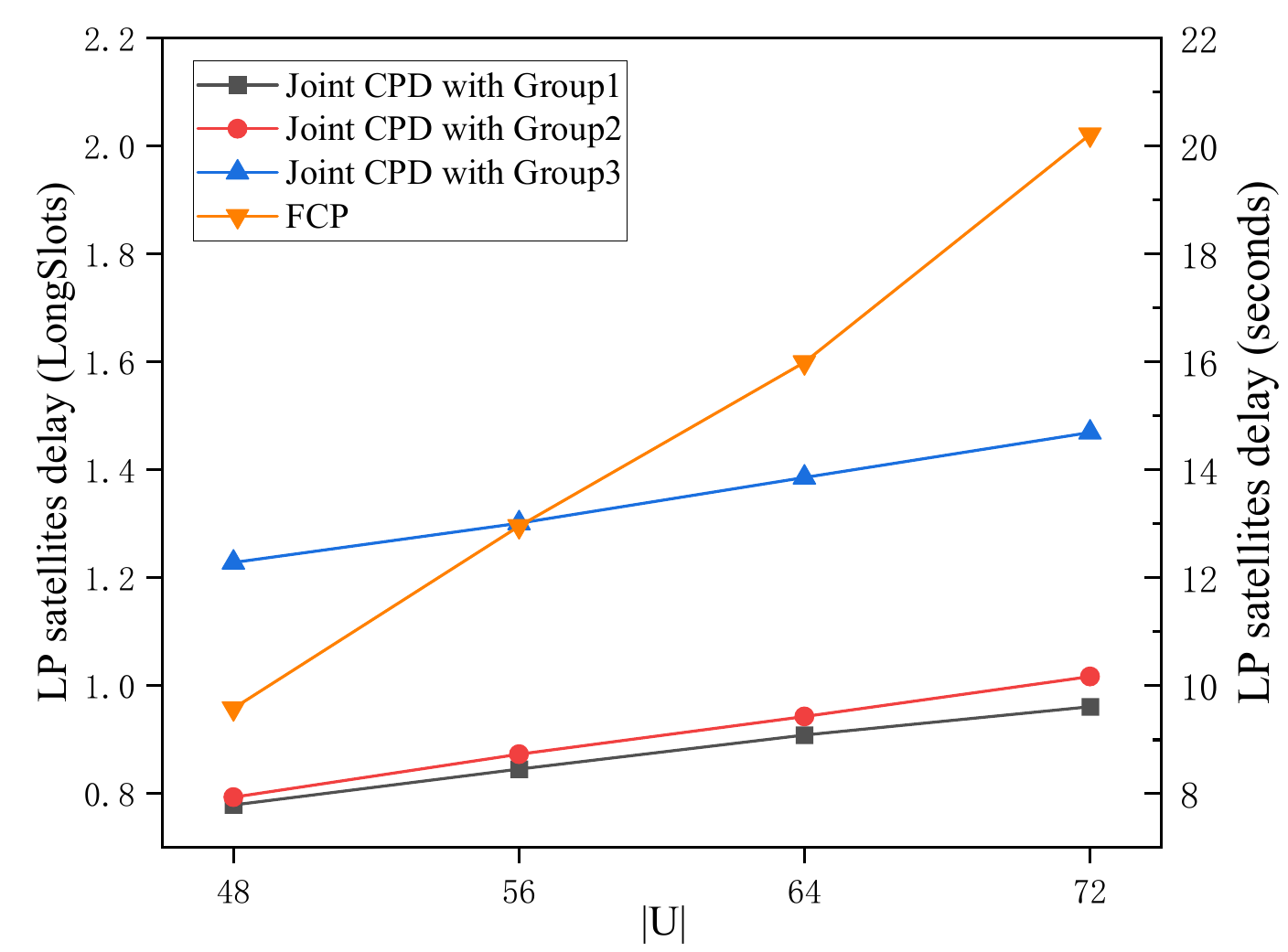}
		\caption{Non-anchor LP satellites to ground station delay.}
		\label{fig3}
	\end{figure}
	
	% Non-Anchor LP Satellites
	\paragraph{Delay for Non-Anchor LP Satellites}
	Fig~\ref{fig3} shows the average delay of non-anchor LP satellites. 
	The overall trend indicates that the delay gradually increases as the number of service users increases, implying that more users consume more communication resources of the LP constellation. 
	% FCP
	The slope of the delay increase for FCP is steeper than that of J-CPD. 
	This is because FCP only considers fairness in establishing links without distinguishing between satellites and users, leading to a more severe performance degradation as the number of users rises. 
	% J-CPD
	In J-CPD, if a non-anchor satellite has not been linked to an anchor satellite for a long time, its communication potential significantly increases, thereby facilitating the establishment of communication links. 
	Under the third group of parameters in J-CPD, the user potential is the highest; hence, during the link scheduling process, the priority of service links is higher than those under the first and second groups of parameters. 
	Consequently, the communication performance under the third group of parameters is worse than that under the first and second groups.
	
	\begin{figure}[] 
		\centering
		\includegraphics[width=0.9\linewidth]{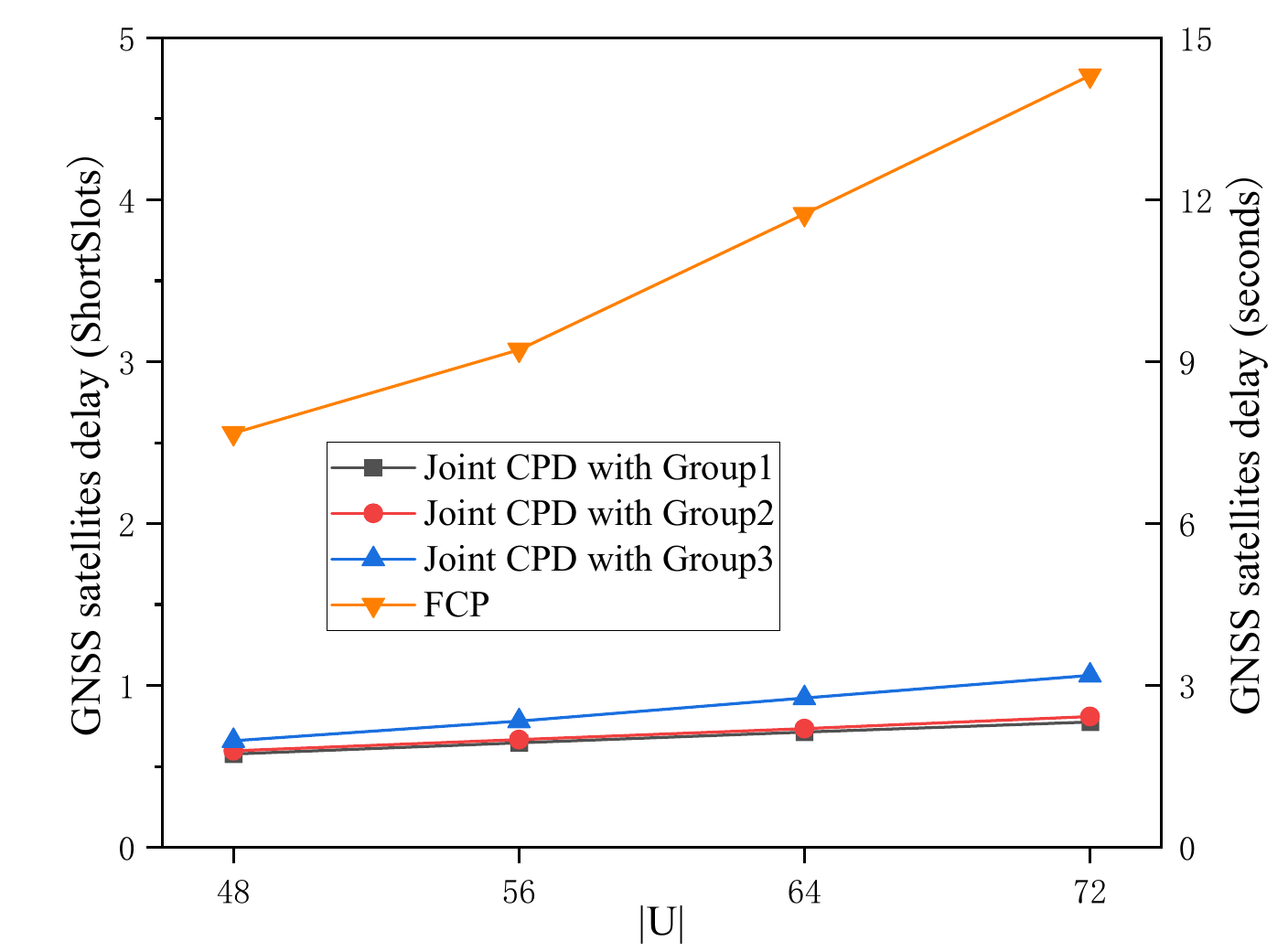}
		\caption{Non-anchor GNSS satellites to ground station delay.}
		\label{fig4}
	\end{figure}
	
	\paragraph{Delay for Non-Anchor GNSS Satellites}
	Fig~\ref{fig4} illustrates the average delay of non-anchor GNSS satellites, which generally follows a similar trend to that of the average delay of non-anchor LP satellites. 
	The difference lies in the gap between the delays under the FCP method and those under the J-CPD method for non-anchor GNSS satellites being larger than that for non-anchor LP satellites. 
	This is because, under the CPD process shown in Fig~\ref{fig2}, we first perform LongSlot CPD. 
	If FCP is used, which only considers fairness in establishing links, a significant portion of the GNSS satellites' resources would be occupied by LP satellites and users. 
	During the ShortSlot CPD, the resources available for internal link establishment within GNSS satellites become limited. 
	However, the exclusion potential proposed for GNSS satellites in J-CPD effectively mitigates this situation, thus resulting in significantly better delay performance compared to FCP.
	
	\subsubsection{Ranging}
	
	Ranging refers to the process of measuring distances between satellites, which is fundamental for accurate positioning, orbit determination, and coordinated communication tasks.  
	This section evaluates the effectiveness of the J-CPD and FCP methods in establishing such links.
	
	\paragraph{Ranging Links for LP Satellites}
	
	\begin{figure}[] 
		\centering
		\includegraphics[width=0.86\linewidth]{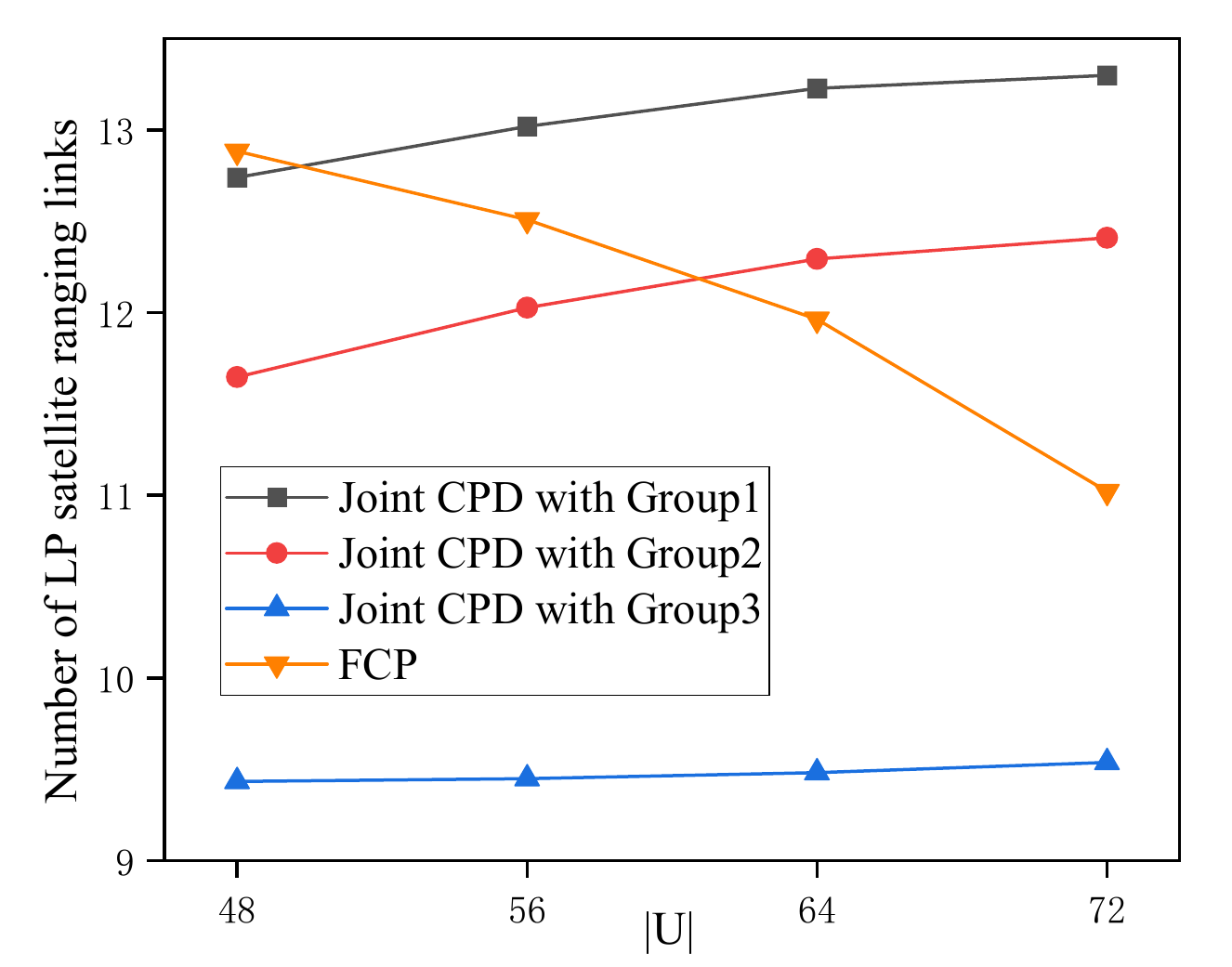}
		\caption{The number of ranging links for LP satellites.}
		\label{fig5}
	\end{figure}
	
	% Fig~\ref{fig5} shows the average number of ranging links obtained by LP satellites. 
	% All methods achieved more ranging links than our expected number of $I_L=5$ links. 
	% However, the number of ranging links under FCP shows a decreasing trend as the number of users increases. This is primarily because FCP does not distinguish between users and satellites; thus, an increase in the number of users consumes more ranging resources between satellites. The fewer ranging links under the third group of parameters in J-CPD compared to the first and second groups are due to the higher user potential, which results in higher priority during link scheduling.
	Fig.~\ref{fig5} illustrates the average number of ranging links established by LP satellites.
	All methods exceed the expected target of $I_L = 5$ links per satellite.
	However, FCP exhibits a downward trend as the number of users increases, since it does not prioritize between users and satellites—more users result in fewer inter-satellite ranging links.
	In contrast, J-CPD maintains better performance, though the third parameter group shows slightly fewer ranging links due to higher user potential, which shifts scheduling priority toward user service over satellite ranging.
	
	\paragraph{Ranging Links for GNSS Satellites}
	
	\begin{figure}[] 
		\centering
		\includegraphics[width=0.86\linewidth]{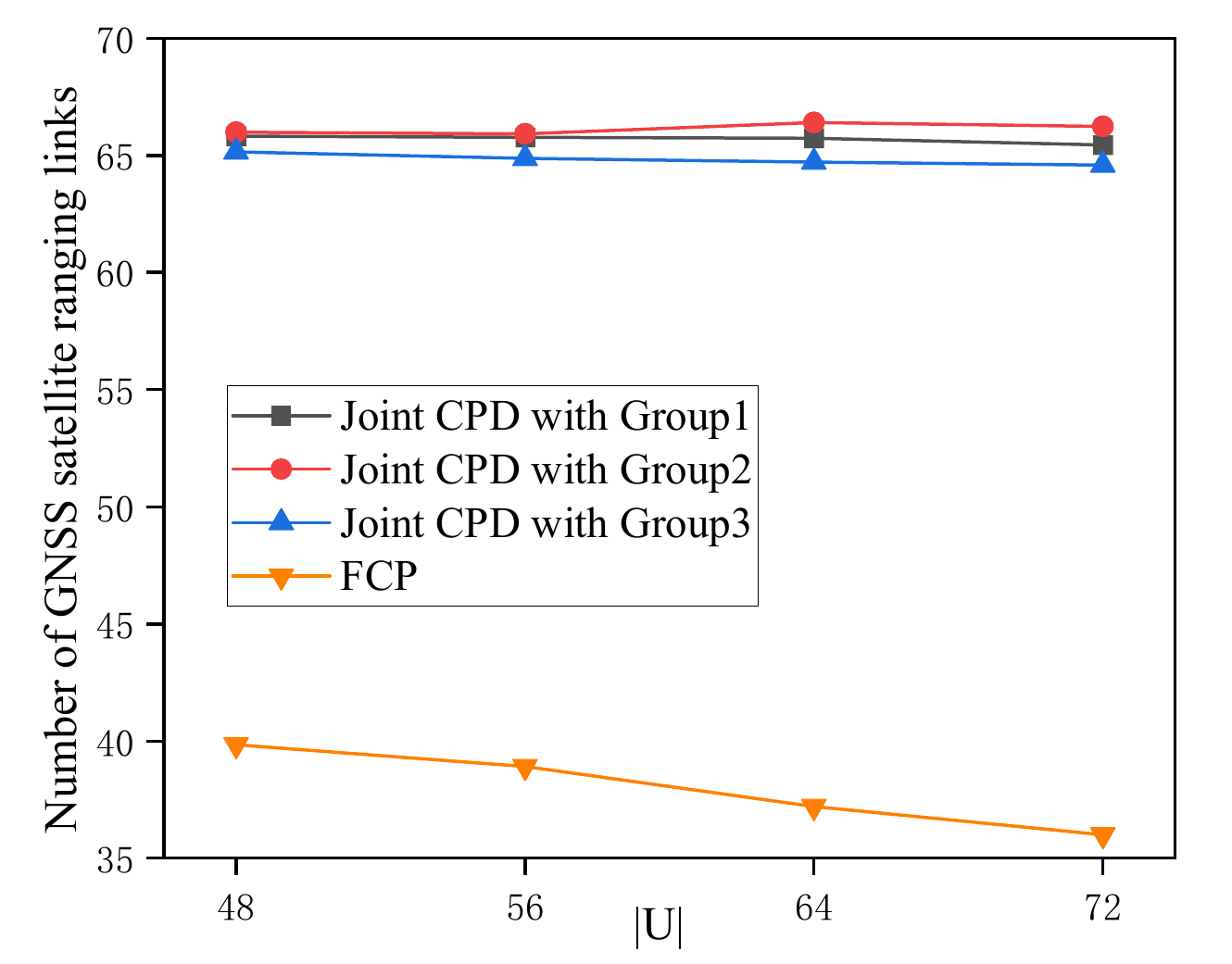}
		\caption{The number of ranging links for GNSS satellites.}
		\label{fig6}
	\end{figure}
	
	% Fig~\ref{fig6} illustrates the average number of ranging links obtained by GNSS satellites. Under the Joint CPD method, GNSS satellites on average acquire more than 60 ranging links. In contrast, under the FCP method, the number of ranging links obtained by GNSS satellites is only around 15, and this number shows a decreasing trend as the number of users increases. The poor performance of FCP can be attributed to two main reasons. 
	% Firstly, during the LongSlot CPD process, LP satellites and users consume a significant portion of the GNSS satellites' resources, leaving limited resources available for GNSS satellites during the ShortSlot CPD. 
	% Secondly, FCP does not consider the necessity of waiting for a certain period before establishing effective ranging links, whereas in the Joint CPD method, ranging potential between satellites only exists after a certain interval. Consequently, during the scheduling process, Joint CPD tends to prioritize the establishment of links that already have ranging potential.
	Fig.~\ref{fig6} presents the average number of ranging links established by GNSS satellites.
	Under J-CPD, each GNSS satellite consistently acquires over 60 links, whereas FCP yields only about 40, with a declining trend as the user count increases.
	FCP underperforms for two main reasons:
	(\textit{i}) In the LongSlot CPD, a large share of GNSS resources is consumed by LP satellites and users, leaving limited capacity for intra-GNSS links during the subsequent ShortSlot CPD.
	(\textit{ii}) FCP lacks awareness of the waiting time required to activate ranging potential. In contrast, J-CPD models this explicitly and prioritizes link scheduling accordingly, favoring links with already established potential.
	
	% Conclusion for Ranging
	The poor performance of FCP in terms of delay and range makes it unsuitable for the joint constellation. 
	In contrast, J-CPD is capable of meeting the communication and ranging requirements of both LP satellites and GNSS satellites in this scenario.

	\begin{table}[]
		\caption{User Satisfaction Ratio}
		\label{user satisfaction ratio}
		\centering
		\centering
		\footnotesize % 调整字体大小以适应表格宽度
		\renewcommand{\arraystretch}{1.2} % 可选：增加行高以提高可读性
		\begin{tabular}{|c|c|c|c|c|}
			\hline
			$|U|$ & \textbf{Group1}& \textbf{Group2}& \textbf{Group3}&\textbf{FCP}  \\ 
			\hline
			48 & 100\%  & 100\%& 100\%  & 100\%  \\
			\hline
			56 & 99.99\%  & 100\% & 100\% & 100\%   \\
			\hline
			64 & 99.87\%  &99.99\%&  100\%  & 100\%  \\
			\hline
			72 & 98.85\%& 99.90\% &  100\%&  100\%  \\
			\hline
		\end{tabular}
	\end{table}
	
	\begin{table*}[]
		\caption{Distribution of Link Types}
		\label{link composition}
		\centering
		\centering
		\footnotesize % 调整字体大小以适应表格宽度
		\renewcommand{\arraystretch}{1.2} % 可选：增加行高以提高可读性
		\begin{tabular}{|c|c|c|c|c|}
			\hline
			\textbf{Link Type} & \textbf{Group1}& \textbf{Group2}& \textbf{Group3}&\textbf{FCP}  \\ 
			\hline
			Number of links between LP satellites and Users & 68.18  & 73.67& 80.58  &36.40   \\
			\hline
			Number of links between GNSS satellites and Users & 187.48  & 182.33 & 175.42 & 220.05   \\
			\hline
			Number of links between LP satellites  & 10.24  &19.72&  20.90  & 0.056  \\
			\hline
			Number of links between GNSS satellites & 1294.03& 1349.03 &  1395.11&  1262.59  \\
			\hline
			Number of links between LP satellites and GNSS satellites& 71.31& 45.56 &  26.87&  123.39  \\
			\hline
		\end{tabular}
	\end{table*}
	\subsubsection{User Satisfaction Ratio}
	% The user satisfaction ratio is the average ratio of the number of ISLs provided by satellites to the number of ISLs required by users in $U$ across all FSA states. 
	The user satisfaction ratio is defined as the average ratio between the number of ISLs successfully provided to each user and the number of ISLs requested, evaluated across all FSA states.

	% Table~\ref{user satisfaction ratio} shows the user satisfaction rates of the J-CPD method and the FCP method under three groups of parameters.
	% We consider that when the user satisfaction rate falls below 100\%, the number of users exceeds the service capacity of the satellites. The service capacities of Group 1 and Group 2 are lower than that of Group 3, as the higher user potential in Group 3 leads to a prioritization of serving users during the link scheduling process. Consequently, the communication and ranging performance in Group 3 are inferior to those in Group 1 and Group 2.
	% Although the user satisfaction rate of FCP also shows relatively good performance, this is achieved at the expense of an unhealthy occupation of the satellites' communication and ranging resources by the users.
	As shown in Table~\ref{user satisfaction ratio}, the J-CPD method achieves near-perfect satisfaction across all scenarios, with Group 3 consistently reaching 100\% regardless of the number of users. 
	When satisfaction drops below 100\%, it indicates that user demand exceeds the system’s service capacity. 
	This occurs in Groups 1 and 2, which assign less weight to user servicing compared to Group 3, where higher user potential ensures users are prioritized during link scheduling.
	
	While the FCP method also maintains full satisfaction, it does so by heavily allocating satellite resources to users, leading to degraded performance in other areas, such as delay and ranging. 
	This highlights a key trade-off: J-CPD achieves more balanced resource allocation across all mission objectives, whereas FCP overcommits to user service at the expense of overall network health.

	\subsubsection{Distribution of Link Types}
	
	Table~\ref{link composition} presents the distribution of established links across five categories under the J-CPD method (for three parameter groups) and the FCP baseline, with $|U|=64$.

	In Group 2, the exclusion potential between LP and GNSS satellites is increased, which substantially reduces the number of inter-constellation links. 
	As a result, LP satellites redirect more resources to internal links and user services, while GNSS satellites focus on intra-constellation connectivity.
	
	In Group 3, the user potential is further amplified. Early in the planning horizon, GNSS satellites retain a high exclusion potential toward users, making LP satellites the primary providers of user service. 
	This leads to more LP-to-user links, while further suppressing LP-GNSS connectivity.
	
	The FCP method, by contrast, does not distinguish between node roles or priorities. 
	It favors proportional allocation, granting more links to the most populous node types. 
	In the early slots, user connections dominate due to their quantity. 
	Once all user demands are met, FCP reallocates resources heavily toward LP-GNSS and intra-GNSS links. 
	This undifferentiated allocation, while maintaining user satisfaction, results in inefficiencies in both communication and link distribution.
	
	\begin{table}[]
		\caption{J-CPD Computation Effort}
		\label{Computation Effort}
		\centering
		\centering
		\footnotesize % 调整字体大小以适应表格宽度
		\renewcommand{\arraystretch}{1.2} % 可选：增加行高以提高可读性
		\begin{tabular}{|c|c|c|c|c|}
			\hline
			& $|U|$=48& $|U|$=56& $|U|$=64& $|U|$=72   \\ 
			\hline
			Group1 & 1011 sec  & 1269 sec & 1427 sec  & 1497 sec \\
			\hline
			Group2 & 1137 sec  & 1287 sec & 1338 sec &1552 sec  \\
			\hline
			Group3 & 1114 sec  &1232 sec &  1276 sec  & 1487 sec  \\
			\hline
			FCP & 692 sec& 751 sec &  800 sec&  972 sec  \\
			\hline
		\end{tabular}
	\end{table}
	
	\subsubsection{Computation Effort}
	Table~\ref{Computation Effort} details computational times for the 7-day simulation scenario executed on an AMD Ryzen 7 7700X (4.5 GHz) workstation with 32 GB RAM. Although J-CPD incurs higher computation time than FCP, its absolute overhead remains negligible, orders of magnitude below the simulated duration, confirming low computational demand. Despite the node count increasing 30\% (81 to 105 nodes) with added users, computation time grows marginally. This demonstrates J-CPD's linear scalability for large-scale constellations.
	
	\subsection{Further Discussion}
	The evaluated simulation results lead to the following key observations:
	
	\begin{enumerate}
		%\item The proposed J-CPD algorithm is capable of meeting the communication and ranging needs of both LP satellites and GNSS satellites while also providing superior external service capabilities. This demonstrates the adaptability of the J-CPD in the joint constellation.
		\item The proposed J-CPD algorithm effectively satisfies both communication and ranging demands for LP and GNSS satellites, while also offering robust user service. This confirms its adaptability to heterogeneous, GNSS-LP multi-layered constellations.
		%\item Due to the specific planning process outlined in this paper, the FCP algorithm is entirely unsuitable for this scenario. The primary reason is that during the LongSlot CPD, users and LP satellites would disproportionately consume the resources of GNSS satellites, leading to significantly degraded ranging and communication performance of the GNSS satellites. This unhealthy resource contention adversely affects the overall system performance.
		\item The legacy FCP algorithm proves inadequate in this scenario. Its fairness-driven strategy fails under the sequential CPD process: during LongSlot CPD, users and LP satellites consume excessive GNSS resources, severely degrading intra-GNSS performance and overall system efficiency.
		%\item This paper demonstrates the performance of the J-CPD method under three groups of parameters. By analyzing the link composition, it is evident how adjusting the parameters specifically impacts link planning. This makes the J-CPD a highly flexible method; based on the system's performance preferences, various solutions can be achieved by tuning the parameters. Consequently, the adaptability and versatility of the J-CPD approach are highlighted, offering tailored solutions to meet diverse performance objectives.
		\item The evaluation across three parameter groups highlights how parameter tuning shapes link planning. J-CPD demonstrates strong flexibility: its parameters can be adjusted to favor different system objectives, making it a versatile tool for mission-specific optimization.
		%\item The methods for adjusting various parameters in J-CPD can be based on personal experience, heuristic algorithms such as simulated annealing, or AI techniques.
		\item Parameter tuning in J-CPD can be guided by expert knowledge, heuristics (e.g., simulated annealing), or data-driven approaches using AI techniques—allowing tailored, performance-aware configurations.
	\end{enumerate}
	
	\subsection{Outlook}
	Future research avenues include:
	\begin{enumerate}
		\item Emerging GNSS architectures are expected to incorporate laser terminals\cite{21,31}, while deep-space optical communication continues to advance\cite{45}. 
		This opens new opportunities for hybrid RF/laser joint constellations—beyond the RF-only scenario considered here. Designing a CPD framework for such integrated systems represents a key direction for future work.
		\item Reference~\cite{48}'s blueprint for cislunar infrastructure integrates terrestrial stations, lunar surface facilities, GNSS constellations, lunar-orbiting constellations, and libration point constellations. The significant orbital heterogeneity across these systems necessitates scheduling solutions for highly heterogeneous link durations. 
		Our J-CPD scheme addresses this via its hierarchical-crossed CPD process and energy-driven heuristic algorithm, providing a standardized methodology. Extending J-CPD to schedule increasingly diverse link durations will support the robust operation of future cislunar infrastructure.
		\item Nearly all existing CPD schemes execute on ground stations, with scheduling results uploaded to satellites. The only distributed CPD scheme proposed in~\cite{25}-while being the first operable on GNSS satellites-lacks the capability for real-time on-board sensing of link status and Doppler shifts to adjust ISL scheduling. In the GNSS-LP scenario addressed in this work, significantly longer link distances and greater orbital variations cause more pronounced link status fluctuations and Doppler shifts compared to GNSS-only constellations. This necessitates developing CPD schemes capable of real-time link status and Doppler shift sensing with dynamic ISL scheduling adaptation.
		\item J-CPD's hierarchical-crossed scheduling principles exhibit synergistic potential with UAV (unmanned aerial vehicle) swarm coordination techniques~\cite{49,50} for next-generation multi-agent systems. Specifically, dual-layer control architectures-where the strategic layer governs long-term objectives (e.g., formation keeping) while the tactical layer handles real-time responses (e.g., collision avoidance)-directly mirror J-CPD's hybrid duration scheduling of LongSlot and ShortSlot ISLs.
		
	\end{enumerate}
	\section{Conclusions}
	\label{sec_conclusion}
	Deploying satellites at libration points (LP) offers a promising path toward building deep-space PNT infrastructure. 
	This work introduces a joint constellation architecture—integrating GNSS and LP satellites—supported by a unified ground segment and designed to provide coordinated communication and ranging services across cislunar space.
	
	To manage the pronounced orbital and temporal heterogeneity between GNSS, LP satellites, and users, we first propose a tailored topology model that incorporates hierarchical link switching via LongSlots and ShortSlots. 
	Based on this model, we develop a two-stage Contact Plan Design (CPD) process: link scheduling over Earth-Moon distances in LongSlots, followed by intra-GNSS link planning in the ShortSlots nested within them.
	
	We then present the Joint CPD (J-CPD) algorithm, inspired by the concept of potential energy. 
	Each satellite and user’s service requirement is modeled as a form of potential energy, and ISL establishment acts to minimize this energy. 
	A maximum weight matching strategy is used to select the most beneficial set of links in each state. 
	Simulation results demonstrate that J-CPD simultaneously supports effective ranging and communication for both GNSS and LP satellites, while ensuring high-quality service for users.
	Moreover, the method is flexible and tunable: adjusting potential parameters allows trade-offs between GNSS performance, LP network efficiency, and user service levels, making J-CPD adaptable to a wide range of mission goals and system constraints.
	To our knowledge, J-CPD constitutes the first CPD framework demonstrated to schedule ISLs with fundamentally heterogeneous durations (ShortSlots vs. LongSlots) in integrated Earth-Moon navigation architectures.

\bibliographystyle{IEEEtran}
\bibliography{references}

\end{document}